\documentclass{LMCS}
 
\usepackage[OT1]{fontenc}
\usepackage{subfigure}

\usepackage{tikz}
\usetikzlibrary{arrows,decorations.pathmorphing,patterns}
\usepackage{amssymb,amsmath}
\usepackage{enumerate}
\usepackage{hyperref}

\makeatletter

\newcommand{\Goal}{\text{\sffamily\upshape{Goal}}}
\newcommand{\oBs}{\text{obs}}

\newcommand\A{{\mathcal A}}
\newcommand\C{{\mathcal C}} 
\newcommand\M{{\mathcal M}}
\renewcommand\P{{\mathcal P}}

\newcommand\Runsf[1]{\text{\sffamily\upshape{Runs}}_{f}(#1)}

\newcommand\Pred{\text{\sffamily\upshape{Pred}}}
\newcommand\timePred{\text{\sffamily\upshape{time-Pred}}}
\newcommand\Post{\text{\sffamily\upshape{Post}}}
\newcommand\cPred{\text{\sffamily\upshape{cPred}}}
\newcommand\uPred{\text{\sffamily\upshape{uPred}}}
\newcommand\last{last}

\newcommand{\Suf}{\text{{\sffamily\upshape{Suf}}}}

\newcommand{\Sup}{\text{{\sffamily\upshape{Sup}}}}
\newcommand{\Time}{\text{{\sffamily\upshape{Time}}}}
\newcommand{\Th}{\text{{\sffamily\upshape{Th}}}}

\newcommand\IN{{\mathbb N}}

\newcommand\IQ{{\mathbb Q}}
\newcommand\IR{{\mathbb R}}

\let\phi=\varphi
\let\epsilon=\varepsilon
\@ifundefined{leqslant}{}{\let\le=\leqslant}
\@ifundefined{geqslant}{}{\let\ge=\geqslant}
\@ifundefined{varnothing}{}{\let\emptyset=\varnothing}
\ifx\upharpoonright\@undefined
  
\else                                                 % $f\restrict E$ or
\fi

\newcommand\ds{{({\mathcal M},\gamma)}}

\newcommand\mkun{V_1}
\newcommand\mkdeux{V_2}
\renewcommand\Pi{{{\mathcal P}_i}}

\let\c@definition\c@theorem
\let\c@lemma\c@theorem
\let\c@corollary\c@theorem
\let\c@remark\c@theorem
\let\c@example\c@theorem
\let\c@proposition\c@theorem
\makeatother

\def\doi{6 (1:1) 2010}
\lmcsheading%
{\doi}
{1--48}
{}
{}
{Apr.~\phantom09, 2007}
{Jan.~12, 2010}
{}   

\begin{document}

\title{O-Minimal Hybrid Reachability Games}

\author[P.~Bouyer]{Patricia Bouyer\rsuper a}
\address{{\lsuper a}LSV, CNRS \& ENS Cachan\\
  61, avenue du Pr{\'e}sident Wilson, 94230 Cachan, France}
\email{\{bouyer,chevalie\}@lsv.ens-cachan.fr}

\author[T.~Brihaye]{Thomas Brihaye\rsuper b}
\address{{\lsuper a}Universit\'e de Mons \\ 
20, place du parc, 7000 Mons, Belgium}
\email{thomas.brihaye@umons.ac.be}

\author{Fabrice Chevalier\rsuper a}

\begin{abstract}
  In this paper, we consider reachability games over general hybrid
  systems, and distinguish between two possible observation frameworks
  for those games: either the precise dynamics of the system is seen
  by the players (this is the perfect observation framework), or only
  the starting point and the delays are known by the players (this is
  the partial observation framework). In the first more classical
  framework, we show that time-abstract bisimulation is not adequate
  for solving this problem, although it is sufficient in the case of
  timed automata.  That is why we consider an other equivalence,
  namely the suffix equivalence based on the encoding of trajectories
  through words. We show that this suffix equivalence is in general a
  correct abstraction for games. We apply this result to o-minimal
  hybrid systems, and get decidability and computability results in
  this framework. For the second framework which assumes a partial
  observation of the dynamics of the system, we propose another
  abstraction, called the superword encoding, which is suitable to
  solve the games under that assumption. In that framework, we also
  provide decidability and computability results.
\end{abstract}


\maketitle

\section{Introduction}

\paragraph{Games over hybrid systems.}
Hybrid systems are finite-state machines equipped with a continuous
dynamics. In the last thirty years, formal verification of such
systems has become a very active field of research in computer
science, with numerous success stories. In this context, hybrid
automata, an extension of timed automata~\cite{AD90,AD94}, have been
intensively studied~\cite{henzinger95,henzinger96}, and decidable
subclasses of hybrid systems have been drawn like initialized
rectangular hybrid automata~\cite{henzinger96}. More recently, games
over hybrid systems have appeared as a new interesting and active
field of research since, among others, they correspond to a
formulation of control problems, the counterpart of model checking for
open systems, \textit{i.e.}, systems embedded in a possibly reactive
environment.  In this context, many results have already been
obtained, like the (un)decidability of control problems for hybrid
automata~\cite{HHM99}, or (semi-)algorithms for solving such
problems~\cite{AHM01}. Given a system $S$ (with controllable and
uncontrollable actions) and a property $\varphi$, controlling the
system means building another system $C$ (which can only enforce
controllable actions), called the controller, such that $S \parallel
C$ (the system $S$ guided by the controller $C$) satisfies the
property $\varphi$. In our context, the property is a reachability
property and our aim is to build a controller enforcing a given
location of the system, whatever the environment does (which plays
with the uncontrollable actions).

\paragraph{O-minimal hybrid systems.} 
O-minimal hybrid systems have been first proposed in~\cite{LPS00} as
an interesting class of systems (see~\cite{vandendries98} for an
overview of properties of o-minimal structures). They have very rich
continuous dynamics, but limited discrete steps (at each discrete
step, all variables have to be reset, independently from their initial
values).  This allows to decouple the continuous and discrete
components of the hybrid system (see \cite{LPS00}). Thus, properties
of a global o-minimal system can be deduced directly from properties
of the continuous parts of the system. Since the introductory
paper~\cite{LPS00}, several works have considered o-minimal hybrid
systems~\cite{davoren99,BMRT04,BM05,KV04,KV06}, mostly focusing on
abstractions of such systems, on reachability properties, and on
bisimulation properties.

\paragraph{Word encoding.}
In~\cite{BMRT04}, an encoding of trajectories with words has been
proposed in order to prove the existence of finite bisimulations for
o-minimal hybrid systems (see also~\cite{BM05}).  Let us mention that
this technique has been used in \cite{KV04,KV06} in order to provide
an exponential bound on the size of the finite bisimulation in the
case of pfaffian hybrid systems. Let us also notice that similar
techniques already appeared in the literature, see for instance the
notion of signature in~\cite{ASY01}. Different word encoding
techniques have been studied in a wider context
in~\cite{brihaye05}. Recently in~\cite{KRS07}, the authors propose a
new algorithm for counter-example guided abstraction and refinement on
hybrid systems, based on use a word encoding approach.  In this paper
we use the so-called \emph{suffix encoding}, which was shown to be in
general too fine to provide the coarsest time-abstract bisimulation.
However, based on this encoding, a semi-algorithm has been proposed
in~\cite{brihaye05,brihaye06} for computing a time-abstract
bisimulation, and it terminates in the case of o-minimal hybrid
systems.

\paragraph{Contributions of this paper.}
In this paper, we focus on games over hybrid systems. We describe two
rather natural frameworks for such games, one assuming a perfect
observation of the dynamics of the system, and another one assuming a
partial observation of the dynamics. For the first framework, we use
the above-mentioned suffix word encoding of trajectories for giving
sufficient computability conditions for the winning states of a
game. Time-abstract bisimulation is an equivalence relation which is
correct with respect to reachability properties on hybrid
systems~\cite{AHLP00} and with respect to control reachability
properties on timed automata~\cite{AMPS98}.  Here, we show that the
time-abstract bisimulation is not correct anymore for solving control
problems on a general class of hybrid systems: we exhibit a system in
which two states are time-abstract bisimilar, but one of the states is
winning and the other is not. Using the suffix encoding of
trajectories of~\cite{brihaye05}, we prove that, in the perfect
observation framework, two states having the same suffixes are
equivalently winning or losing (this is a stronger condition than the
one for the time-abstract bisimulation). We then focus on o-minimal
hybrid games and prove that, under the assumption that the theory of
the underlying o-minimal structure is decidable, the control problem
can be solved and that winning states and winning strategies can be
computed. Regarding the partial observation framework, we provide a
new encoding technique, the so-called superword encoding, which turns
out to be sound for the control under partial observation of the
dynamics, and which allows to prove decidability and computability
results similar to those in the perfect observation framework.

\paragraph{Related work.}
The most relevant related works are those dealing with hybrid
games~\cite{HHM99,AHM01}.  However, the framework of these papers is
pretty different from ours:
\begin{enumerate}
\item
In their framework, time is considered as a
discrete action, and once action ``let time elapse'' has been chosen,
it is not possible to bound the time elapsing, which is quite
restrictive. For instance, the timed game of Figure~\ref{fig:bete} is
winning from $(\ell_0,x=0)$ in our framework (the strategy is to wait
some amount of time $t \in [2,5]$ and to take the controllable action
$c$), whereas it is not winning in their framework (once $x$ is above
$5$, it is no more possible to take the transition and reach the
winning location $\ell_1$, and there is no way to impose a delay
within $[2,5]$). This yields significant differences in the
properties: in their framework, game bisimulation is one of the tools
for solving the games, and as stated by~\cite[Prop.~1]{HHM99}, the
classical bisimulation tool is then sufficient to solve games. On the
contrary, in our framework, the notion of bisimulation relevant to our
model (time-abstract bisimulation) is not correct for solving games,
as will be explored in this paper.
\begin{figure}[ht]
  \centering
  \begin{tikzpicture}
    \draw (0,0) node [draw,circle,inner sep=1.5pt] (A) {$\ell_0$};
    \draw (3,0) node [draw,circle,inner sep=1.5pt,fill=black!40!white] (B) {$\ell_1$};
    \draw [latex'-] (A) -- +(-.8,0);
    \draw [-latex'] (A) -- (B) node [midway,above] {$2 \le x \le 5,\ c$};
  \end{tikzpicture}
  \caption{A simple game}
  \label{fig:bete}
\end{figure}
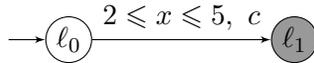

\item Our games are control games, they are thus asymmetric, which is
  not the case of the games in the above-mentioned works; in our
  framework, the environment is more powerful than the controller in
  that it can outstrip the controller and do an action right before
  the controller decides to do a controllable action.
\end{enumerate}
Let us also mention the paper~\cite{WT97} on control of linear hybrid
automata.  In~\cite{WT97} the author proposes a semidecision procedure
for synthesizing controllers for such automata. No general
decidability result is given in this paper.

\paragraph{Plan of the paper.}
In Section~\ref{untimed}, we recall results about finite games and
bisimulation.  In Section~\ref{timed}, we define the games over
dynamical systems (for both perfect information and partial
observation), and we show that time-abstract bisimulation is not
correct for solving them. The word encoding techniques are presented
in Section~\ref{section2} and used in Section~\ref{mgame} to present a
general framework for solving games over dynamical systems. We apply
and extend these results in Section~\ref{omingames} for computing
winning states and winning strategies in o-minimal games. In the
paper, we often only develop technical details of the partial
observation framework, which actually extends the perfect observation
framework.

\bigskip Part of the results presented in this paper have been
published in~\cite{BBC06} (the decidability of the control
reachability problem and the synthesis of strategies for o-minimal
hybrid systems).  In this paper, we give full proofs of those results,
and extend them to a natural partial observation framework.

\section{Classical Finite Games}\label{untimed}

In this section, we recall some basic definitions and results
concerning bisimulations on a transition system (see
\cite{aczel88,milner89,caucal95,henzinger95} for general references)
and classical (untimed) games.

\subsection{Classical Games}

We present here the definitions of the problem of control on a finite
graph (also called finite game) and the notion of strategy
(see~\cite{lncs2500} for an overview on games). These definitions are
classical and will be extended to real-time systems in the next
section.

\begin{defi}
  A \emph{finite automaton} is a tuple $\A=(Q,\Goal,\Sigma,\delta)$
  where $Q$ is a finite set of locations, $\Goal \subseteq Q$ is a
  subset of winning locations, $\Sigma$ is a finite set of actions,
  and $\delta$ consists of a finite number of transitions $(q,a,q')
  \in Q \times \Sigma \times Q$.
\end{defi}

\begin{defi} \label{trans syst def} A \emph{transition system}
  $T=(Q,\Sigma,{\to})$ consists of a set of states $Q$ (which may be
  uncountable), $\Sigma$ an alphabet of events, and ${\to} \subseteq Q
  \times \Sigma \times Q$ a transition relation.
\end{defi}

A transition $(q_1,a,q_2) \in {\to}$ is also denoted by $q_1
\xrightarrow{a} q_2$. A transition system is said finite if $Q$ is
finite. Note that a finite automaton canonically defines a transition
system $T_{\A}$.
  
\medskip A {\em run} of $\A$ is a finite or infinite sequence $q_0
\xrightarrow{a_1} q_1 \xrightarrow{a_2} \ldots$ of the transition
system $T_{\A}$. Such a run is said {\em winning} if $q_i \in \Goal$
for some $i$.  If $\rho$ is a finite run $q_0 \xrightarrow{a_1} q_1
\xrightarrow{a_2} \ldots \xrightarrow{a_n} q_n$ we define
$\last(\rho)=q_n$.  We note $\Runsf{\A}$ the set of finite runs in
$\A$.

\begin{defi}
  A {\em finite game} is a finite automaton $(Q,\Goal,\Sigma,\delta)$
  where $\Sigma$ is partitioned into two subsets $\Sigma_c$ and
  $\Sigma_u$ corresponding to controllable and uncontrollable actions.
\end{defi}

We will consider \emph{control games}. Informally there are two
players in such a game: the \emph{controller} and the
\emph{environment}. The actions of $\Sigma_c$ belong to the controller
and the actions of $\Sigma_u$ belong to the environment. At each step,
the controller proposes a controllable action which corresponds to the
action he wants to perform; then either this action or an
uncontrollable action is done and the automaton goes into one of the
next states\footnote{There may be several next states as the game is
  not supposed to be deterministic, and we assume that the environment
  chooses the next state in case there are several.}.
In the sequel, we will only consider reachability games : the
controller wants to reach the $\Goal$ states and the environment wants
to prevent him from doing so.

\begin{defi}
  A {\em strategy} is a partial function $\lambda$ from $\Runsf{\A}$
  to $\Sigma_c$ such that for all runs $\rho \in \Runsf{\A}$, if
  $\lambda(\rho)$ is defined, then it is enabled in $\last(\rho)$.
\end{defi}

Let $\rho = q_0 \xrightarrow{a_1} q_1 \xrightarrow{a_2} \ldots$ be a
run, and set for every $i$, $\rho_i$ the prefix of length $i$ of
$\rho$. The run $\rho$ is said \emph{compatible with a strategy
  $\lambda$} when for all $i$, $a_{i+1}=\lambda(\rho_i)$ or
$a_{i+1}\in \Sigma_u$. A run $\rho$ is said \emph{maximal w.r.t. a
  strategy $\lambda$} if it is infinite or if $\lambda(\rho)$ is not
defined.

A strategy $\lambda$ is \emph{winning from a state q} if all maximal
runs starting in $q$ compatible with $\lambda$ are winning.

\subsection{Bisimulation}

We recall now the definition of bisimulation for transition systems:

\begin{defi}[\cite{milner89,caucal95}]
  Given a transition system $T=(Q,\Sigma,{\to})$, a {\em bisimulation
    for $T$} is an equivalence relation $\mathord\sim \subseteq Q
  \times Q$ such that $\forall q_1,q'_1,q_2 \in Q$, $\forall a \in
  \Sigma$,
  \[\begin{array}{l} \left( q_1 \sim q'_1\ 
      \text{and}\ q_1 \xrightarrow{a} q_2 \right) \Rightarrow \left(
      \exists q'_2\ q_2 \sim q'_2\ \text{ and }\ q'_1
      \xrightarrow{a} q'_2 \right) \qquad \\
  \end{array}\]
  Moreover, if $\P$ is a partition of $Q$ and if $\sim$ respects $\P$
  (\textit{i.e.}, $q \in P$  and  $q \sim q'$ with $P \in \P$ implies $q'
  \in P$), we say that $\sim$ is \emph{compatible} with $\P$. 
\end{defi}

\subsection{Game and Bisimulation in the Untimed Case}

In the untimed framework, bisimulation is a commonly used technique to
abstract games: bisimilar states can be identified in the control
problem. This is stated in the next folklore theorem, for which we
provide a proof.

\begin{thm}
  Let $\A=(Q,\Goal,\Sigma,\delta)$ be a finite game, $q,q' \in Q$ and
  $\mathord\sim$ a bisimulation compatible with $\Goal$. Then, there
  is a winning strategy from $q$ iff there is a winning strategy from
  $q'$.
\end{thm}

\proof
  Assume that $\sim$ is a bisimulation relation compatible with
  $\Goal$ and such that $q \mathrel\sim q'$.  Assume furthermore that
  $\lambda$ is a winning strategy from $q$. We will define a strategy
  $\lambda'$ that will be winning from $q'$. To do that we will map
  finite runs starting in $q'$ to finite runs starting in $q$, so that
  $\lambda'$ will mimick $\lambda$ through this mapping. We note $f$
  this mapping, and start by setting $f(q') = q$. We then proceed
  inductively as follows.  If $\lambda(f(\varrho'))$ is defined, we
  set $\lambda'(\varrho') = \lambda(f(\varrho'))$ and for every run
  $\varrho' \xrightarrow{\lambda'(\varrho'))} \widetilde{q}'$ (which
  is compatible with $\lambda'$) there is a run $f(\varrho')
  \xrightarrow{\lambda(\varrho)} \widetilde{q}$ which is compatible
  with $\lambda$ and such that $\widetilde{q} \mathrel\sim
  \widetilde{q}'$. We then define $f(\varrho'
  \xrightarrow{\lambda'(\varrho')} \widetilde{q}') = f(\varrho')
  \xrightarrow{\lambda(\varrho)} \widetilde{q}$. The strategy
  $\lambda'$ is winning from $q'$ since $\mathord\sim$ is compatible
  with $\Goal$.  \qed

This theorem remains true for infinite-state discrete
games~\cite{HHM99,AHM01} and can be used to solve them: if an
infinite-state game has a bisimulation of finite index, the control
problem can be reduced to a control problem over a finite graph.
Real-time control problems cannot be seen as classical infinite-state
games because of the special nature of the time-elapsing action.
which does not belong to one of the players.  It seems nevertheless
natural to try to adapt the bisimulation approach to solve real-time
control problems.

\section{Games over Dynamical Systems}\label{timed}

\subsection{Dynamical Systems}\label{dynamics}

Let $\M$ be a structure. When we say that some relation, subset or
function is \emph{definable}, we mean it is first-order definable in
the structure $\M$. A general reference for first-order logic is
\cite{hodges97}. We denote by $\Th(\M)$ the theory of $\M$.  In this
paper we only consider structures $\M$ that are expansions of ordered
groups, we also assume that the structure $\M$ contains two symbols of
constants, {\it i.e.}, $\M = \langle M,+,0,1,<,\ldots \rangle$ where
$+$ is the group operation and w.l.o.g. we assume that $0<1$.

\begin{defi}\label{defi ds}
  A \emph{dynamical system} is a pair $(\M,\gamma)$ where:
  \begin{itemize}
  \item $\M = \langle M, +,0,1,<,\ldots \rangle$ is an expansion of an
    ordered group,
  \item $\gamma: \mkun \times V \to \mkdeux$ is a function definable
    in $\M$
    (where $V_1 \subseteq M^{k_1}$, $V \subseteq M$ and $V_2 \subseteq
    M^{k_2}$).\footnote{We use these notations in the rest of the
      paper.}
  \end{itemize}
  The function $\gamma$ is called the \textit{dynamics} of the system.
\end{defi}

Classically, when $M$ is the field of the reals, we see $V$ as the
time, $V_1$ as the input space, $V_1 \times V$ as the space-time and
$V_2$ as the (output) space. We keep this terminology in the more
general context of a structure $\M$.

\medskip The definition of \emph{dynamical system} encompasses a lot
of different behaviors. Let us first give a simple example, several
others will be presented later.

\begin{exa} \label{ex:AT} We can recover the continuous dynamics
  of \emph{timed automata} (see \cite{AD94}). In this case, we have
  that $\M = \langle\IR,<,+,0,1 \rangle$ and the dynamics $\gamma:
  \IR^n \times [0,+\infty[ \to \IR^n$ is defined by
  $\gamma(x_1,\ldots,x_n,t) = (x_1+t,\ldots,x_n+t)$.
\end{exa}

\begin{defi}\label{trajectory}
  If we fix a point $x \in \mkun$, the set $\Gamma_x=\{\gamma(x,t)
  \mid t \in M^+\} \subseteq \mkdeux$ is called the {\em trajectory}
  determined by $x$.
\end{defi}

We define a transition system associated with the dynamical system.
This definition is an adaptation to our context of the classical
\emph{continuous transition system} in the case of hybrid systems (see
\cite{LPS00} for example).

\begin{defi}\label{tsds}
  Given $\ds$ a dynamical system, we define a \emph{transition system
    $T_\gamma=(Q,\Sigma,\to_\gamma)$ associated with the dynamical
    system} by:
  \begin{itemize}
  \item the set $Q$ of states is $\mkdeux$;
  \item the set $\Sigma$ of events is $M^+=\{ \tau \in M \mid \tau \ge
    0\}$;
  \item the transition relation $y_1 \xrightarrow{t}_\gamma y_2$ is
    defined by:
    \begin{align*}
      &\exists x \in \mkun,\ \exists t_1, t_2 \in M^+\ \text{such that
      } t_1 \le t_2,\\
      &\qquad \gamma(x,t_1)=y_1,\ \gamma(x,t_2)=y_2\ \text{and}\ t =
      t_2-t_1
    \end{align*}
  \end{itemize}
\end{defi}

\subsection{\texorpdfstring{$\M$}{M}-Games Under Perfect 
   Observation}\label{subsec-perfect}
In this subsection, we define $\M$-automata, which are automata with
guards, resets and continuous dynamics definable in the
$\M$-structure.  We then introduce our model of dynamical game which
is an $\M$-automaton with two sets of actions, one for each player; we
finally express in terms of winning strategy the main problem we will
be interested in, the control problem in a class $\C$ of $\M$-automata
under perfect observation. The partial observation framework will be
discussed in Subsection~\ref{subsec-partial}.

\begin{defi}[$\M$-automaton]
  An \emph{$\M$-automaton} $\A$ is a tuple $(\M, Q,\Goal,
  \Sigma,\delta,\gamma)$ where $\M= \langle M,+,0,1,<,\ldots \rangle$
  is an expansion of an ordered group, $Q$ is a finite set of
  locations, $\Goal \subseteq Q$ is a subset of winning locations,
  $\Sigma$ is a finite set of actions, $\delta$ consists in a finite
  number of transitions $(q,g,a,R,q') \in Q \times 2^{V_2} \times
  \Sigma \times (V_2 \to 2^{V_2}) \times Q$ where $g$ and $R$ are
  definable in $\M$, and $\gamma$ maps every location $q\in Q$ to a
  dynamics $\gamma_q: V_1 \times V \to V_2$.
\end{defi}

We use a general definition for resets: a reset $R$ is indeed a
general function from $\mkdeux$ to $2^{\mkdeux}$, which may correspond
to a non-deterministic update. If the current state is $(q,y)$ the
system will jump to some $(q',y')$ with $y' \in R(y)$.

\medskip An $\M$-automaton $\A=(\M,Q,\Goal,\Sigma,\delta,\gamma)$
defines a {\em mixed transition system} $T_{\A}=(S,\Gamma,\to)$ where:
\begin{itemize}
\item the set $S$ of states is $Q \times \mkdeux$;
\item the set $\Gamma$ of labels is $M^+ \cup \Sigma$, (where $M^+=\{
  \tau \in M\ \mid\ \tau \ge 0\}$); 
\item the transition relation $(q,y) \xrightarrow{e} (q',y')$ is
  defined when:
  \begin{itemize}
  \item $e \in \Sigma$, and there exists $(q,g,e,R,q') \in \delta$
    with $y \in g$ and $y' \in R(y)$, or
  \item $e \in M^+$, $q=q'$, and $y \xrightarrow{e}_{\gamma_q} y'$
    where $\gamma_q$ is the dynamic in location $q$.
  \end{itemize}
\end{itemize}

In the sequel, we will focus on behaviors of $\M$-automata which
alternate between continuous transitions and discrete transitions.

We will also need more precise notions of transitions. When $(q,y)
\xrightarrow{\tau} (q,y')$ with $\tau \in M^+$, this is due to some
choice of $(x,t) \in V_1 \times V$ such that $\gamma_q(x,t)=y$. We say
that $(q,y) \xrightarrow{\tau}_{x,t} (q,y')$ if $\gamma_q(x,t)=y$ and
$\gamma_q(x,t+\tau)=y'$. To ease the reading of the paper, we will
sometimes write $(q,x,t,y) \xrightarrow{\tau} (q,x,t+\tau,y')$ for
$(q,y) \xrightarrow{\tau}_{x,t} (q,y')$.  We say that an action
$(\tau,a) \in M^+ \times \Sigma$ is enabled in a state $(q,x,t,y)$ if
there exists $(q',x',t',y')$ and $(q'',x'',t'',y'')$ such that
$(q,x,t,y) \xrightarrow{\tau} (q',x',t',y') \xrightarrow{a}
(q'',x'',t'',y'')$. We then write $(q,x,t,y) \xrightarrow{\tau,a}
(q'',x'',t'',y'')$.

A {\em run} of $\A$ is a finite or infinite sequence
$(q_0,x_0,t_0,y_0) \xrightarrow{\tau_1,a_1} (q_1,x_1,t_1,y_1) \ldots$
Such a run is said {\em winning} if $q_i \in \Goal$ for some $i$.

We note $\Runsf{\A}$ the set of finite runs in $\A$.  If $\rho$ is a
finite run $(q_0,x_0,t_0,y_0) \xrightarrow{\tau_1,a_1} \ldots
\xrightarrow{\tau_n,a_n} (q_n,x_n,t_n,y_n)$ we define $\last(\rho) =
(q_n,x_n,t_n,y_n)$.

\begin{defi}[$\M$-game]
  An {\em $\M$-game} is an $\M$-automaton $(\M,Q,\Goal,\Sigma,$
  $\delta,\gamma)$ where $\Sigma$ is partitioned into two subsets
  $\Sigma_c$ and $\Sigma_u$ corresponding to controllable and
  uncontrollable actions.
\end{defi}

\begin{defi}[Strategy]
  A {\em strategy}\footnote{In the context of control problems, a
    strategy is also called a {\em controller}.} is a partial function
  $\lambda$ from $\Runsf{\A}$ to $M^+ \times \Sigma_c$ such that for
  all runs $\rho$ in $\Runsf{\A}$, if $\lambda(\rho)$ is defined, then
  it is enabled in $\last(\rho)$.
\end{defi}

The strategy tells what is to be done at the current moment: at each
instant it tells what delay we will wait and which controllable action
will be taken after this delay. Note that the environment may have to
choose between several edges, each labeled by the action given by the
strategy (because the original game is not supposed to be
deterministic).

A strategy $\lambda$ is said \emph{memoryless} if for all finite runs
$\rho$ and $\rho'$, $\last(\rho)=\last(\rho')$ implies
$\lambda(\rho)=\lambda(\rho')$.  Let $\rho = (q_0,x_0,t_0,y_0)
\xrightarrow{\tau_1,a_1} \ldots$ be a run, and set for every $i$,
$\rho_i$ the prefix of length $i$ of $\rho$. The run $\rho$ is said
\emph{consistent with a strategy $\lambda$} when for all $i$, if
$\lambda(\rho_i) = (\tau,a)$ then either $\tau_{i+1}=\tau$ and
$a_{i+1}=a$, or $\tau_{i+1} \le \tau$ and $a_{i+1} \in \Sigma_u$.  A
run $\rho$ is said \emph{maximal w.r.t. a strategy $\lambda$} if it is
infinite or if $\lambda(\rho)$ is not defined. 
A strategy $\lambda$ is \emph{winning from a state (q,y)} if for all
$(x,t)$ such that $\gamma(x,t)=y$, all maximal runs starting in
$(q,x,t,y)$ compatible with $\lambda$ are winning.  The \emph{set of
  winning states} is the set of states from which there is a winning
strategy.

\medskip We can now define the control problems we will study.

\begin{prob}[Control problem under perfect observation in a class
  $\C$ of $\M$-automata]
  \label{prob:controlperfect}
  Given an $\M$-game $\A \in \C$, and a definable initial state
  $(q,y)$, determine whether there exists a winning strategy
  in $\A$ from $(q,y)$.
\end{prob}

\begin{prob}[Controller synthesis under perfect observation in a
  class $\C$ of $\M$-automata]
  \label{prob:synthperfect}
  Given an $\M$-game $\A \in \C$, and a definable initial state
  $(q,y)$, determine whether there exists a winning strategy, and
  compute such a strategy if possible.\footnote{In this definition,
    `compute a strategy' means `give a formula for the strategy'. In
    particular, a strategy which is computable is definable in the
    theory.}
\end{prob}

\begin{exa}\label{ex:spiraletot}
  Let us consider the $\M$-game $\A =
  (\M,Q,\Goal,\Sigma,\delta,\gamma)$ (depicted in
  Fig.~\ref{fig:ex-spiraltot}) where ${\mathcal M} = \langle
  \IR,+,\cdot,0,1,<,\sin, \cos \rangle$, $Q=\{q_1,q_2,q_3\}$,
  $\Goal=\{q_2\}$, $\Sigma=\Sigma_c \cup \Sigma_u$ where
  $\Sigma_c=\{c\}$ (resp. $\Sigma_u=\{u\}$) is the set of controllable
  (resp. uncontrollable) actions. The dynamics in $q_1$,
  $\gamma_{q_1}:\IR^2 \times [0,2 \pi] \times
  \IR \to \IR^2$ is defined as follows. \\
  $$\gamma_{q_1}(x_1,x_2,\theta,t)=
  \begin{cases}
    (t.\cos(\theta), t.\sin(\theta)) & \text{ if } (x_1,x_2) = (0,0),\\
    (x_1+t.x_1 , x_2 +t.x_2) & \text{ if } (x_1,x_2) \ne (0,0).
  \end{cases}$$ We associate with this dynamical system the partition
  $\P=\{A,B,C\}$ where $A=\{(0,0)\}$, $B=\{\big(\theta
  \cos(\theta),\theta \sin(\theta)\big) \mid 0 < \theta \le 2 \pi\}$
  and $C=\IR^2 \setminus (A \cup C)$. Let us call piece $B$ \emph{the
    spiral} (see Figure~\ref{fig:spirale}). The guard $g_B$
  corresponds to $B$-states ({\it i.e.}, points on the spiral) and the
  guard $g_C$ corresponds to $C$-states (points not on the spiral and
  different from the origin).
  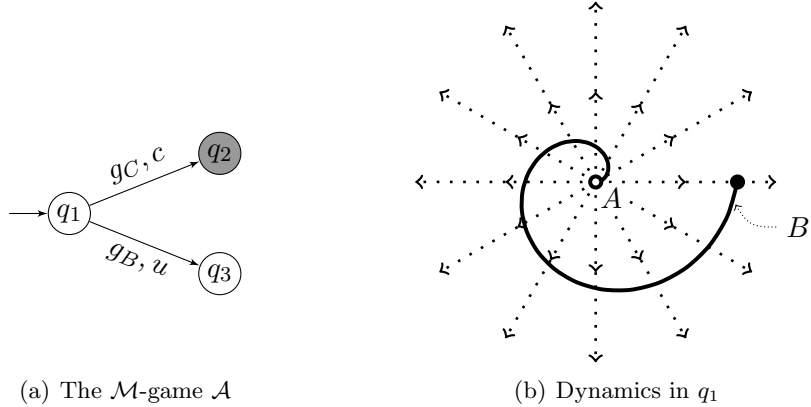
\begin{figure}[ht]
    \null\hfill\hfill\subfigure[The $\M$-game $\A$\label{fig:mgame}]{
      \begin{tikzpicture}
        \draw (0,0) node [circle,inner sep=1.5pt,draw] (q1) {$q_1$};
        \draw [latex'-] (q1) -- +(-.8,0);
        \draw (2,.8) node [circle,inner sep=1.5pt,draw,fill=black!40!white] (q2) {$q_2$};
        \draw (2,-.8) node [circle,inner sep=1.5pt,draw] (q3) {$q_3$};
        \draw [-latex'] (q1) -- (q2) node [midway,sloped,above] {$g_C,c$};
        \draw [-latex'] (q1) -- (q3) node [midway,sloped,below] {$g_B,u$};
        
        \path (1,-2) --(1,-2);
      \end{tikzpicture}
    }\hfill\hfill
    \subfigure[Dynamics in $q_1$\label{fig:spirale}]{
    \begin{tikzpicture}[scale=.3]
      \draw[domain=0:6.283,smooth,variable=\t,line width=1.5pt] plot ({\t*cos(\t r)},{\t*sin(\t r)});
      \foreach \theta in {0,0.524,1.047,1.571,2.094,2.618,3.141,3.665,4.189,4.712,5.236,5.760}
      {
        \draw[domain=0:4,variable=\t,loosely dotted,->,line width=1pt] plot ({\t*cos(\theta r)},{\t*sin(\theta r)});
        \draw[domain=4:8,variable=\t,loosely dotted,->,line width=1pt] plot ({\t*cos(\theta r)},{\t*sin(\theta r)});
      }
      \draw (0,0) [fill=white,line width=1.5pt] circle (7pt) node [below right=1.5pt,fill=white,circle,inner sep=0pt] {$A$}; 
      \draw (6.283,0) [fill=black,line width=1.5pt] circle (7pt); 
      \draw [<-,densely dotted] (6.2,-1) .. controls +(-60:30pt) and +(180:30pt) .. (8,-2) node [pos=1,right] {$B$};
    \end{tikzpicture} 
  }
  \hfill\hfill\null
    \caption{Time-abstract bisimulation does not preserve winning states}
    \label{fig:ex-spiraltot}
  \end{figure}
  In this example, the point $(q_1,(0,0))$ is a winning state. Indeed
  a winning strategy is given by $\lambda(q_1,0,0,\theta,t) =
  (\frac{\theta}{2},c)$ where $c$ consists in taking the transition
  leading to state $q_2$ (which is winning).
\end{exa}

\subsection{\texorpdfstring{$\M$}{M}-Games Under Partial  Observation}\label{subsec-partial}
Subsection~\ref{subsec-perfect}, we have assumed that from a given
point, the environment chooses the continuous trajectory followed by
the game, and the controller reacts accordingly. In this section, we
consider partial observation of the dynamics: the trajectory is not
known by the controller, and its strategy may depend only on the
current point. In particular, this framework naturally models drift of
clocks where the slopes of the clocks lies within an
interval~\cite{puri98,ALM05}.  Note that our partial observation
assumption concerns the dynamics of the system, not the actions which
are performed. This has to be contrasted with the notion of partial
observation studied in the framework of finite systems in~\cite{AVW03}
or in the context of timed systems in~\cite{BDMP03} where the partial
observation assumption concerns actions which are done, and not the
dynamics (indeed, in these models, there is no real choice for the
dynamics; It is completely determined by the point in the
state-space). In order to formalize our partial observation framework,
we need to adapt notions such as strategy in this new setting. First,
we define what we call \emph{observation} of a given run.

\begin{defi}[Observation of a run]
Let $\rho=(q_0,x_0,t_0,y_0)\xrightarrow{\!\tau_1,a_1\!} \ldots
\xrightarrow{\!\tau_n,a_n\!} (q_n,x_n,t_n,y_n)$ be a finite run. The
\emph{observation} of $\rho$, denoted $\oBs(\rho)$ is the sequence
$(q_0,y_0) \xrightarrow{\tau_1,a_1} \ldots \xrightarrow{\tau_n,a_n}
(q_n,y_n)$.
\end{defi}

\begin{defi}[Strategy under partial observation]
  A strategy $\lambda$ is said \emph{under partial observation} if for
  all finite runs $\rho, \rho'$, $\oBs(\rho)=\oBs(\rho')$ implies
  $\lambda(\rho)=\lambda(\rho')$.
\end{defi}

All other notions, like memoryless strategies, consistency, winning
strategies, winning states, \textit{etc...} naturally extend in this
new context. In this setting, we will consider the two following
problems. 

\begin{prob}[Control problem under partial observation in a class
  $\C$ of $\M$-automata]
  \label{prob:contpartial}
  Given an $\M$-game $\A \in \C$, and a definable initial state
  $(q,y)$, determine whether there exists a winning strategy under
  partial observation in $\A$ from $(q,y)$.
\end{prob}

\begin{prob}[Controller synthesis under partial observation in a
  class $\C$ of $\M$-automata]
  \label{prob:synthpartial}
  Given an $\M$-game $\A \in \C$, and a definable initial state
  $(q,y)$, determine whether there exists a winning strategy under
  partial observation in $\A$ from $(q,y)$, and compute such a
  strategy if possible.
\end{prob}

\begin{exa}
  We consider again the spiral example (Example~\ref{ex:spiraletot}).
  We showed that under perfect observation this $\M$-game has a
  winning strategy in $(q_1,(0,0))$ given by
  $\lambda(q_1,0,0,\theta,t) = (\frac{\theta}{2},c)$. Note that this
  strategy depends on the precise trajectory (parameter $\theta$).
  Moreover, one can show that there is no winning strategy under
  partial observation for this game: such a strategy may only depend
  on the current point, and in this precise example, whatever action
  $(\tau,a)$ the controller proposes in $(q_1,(0,0))$, there is a
  trajectory which reaches a \emph{bad} state (\textit{i.e.}, points
  on the spiral) before $\tau$.
\end{exa}

The previous example shows that some games can be winning under
perfect observation whereas they are not winning under partial
observation.  Nevertheless, considering a new dynamics which will
roughly inform the controller of the current trajectory, we can see
the perfect observation control problem as a special case of the
partial observation framework. This is stated by the following
proposition :

\begin{prob} \label{prop:casparticulier} Given an $\M$-game
  $\A_1$ and a state $(q,y)$ of $\A_1$,
  we can effectively construct an $\M$-game $\A_2$ and a state
  $(q',y')$ of $\A_2$ such that there exists a winning strategy under
  perfect observation in $\A_1$ from $(q,y)$ iff there exists a
  winning strategy under partial observation in $\A_2$ from
  $(q',y')$.
\end{prob}
\proof
  Let $\A_1=(\M, Q,\Goal, \Sigma,\delta,\gamma)$ where $\gamma:V_1
  \times V \to V_2$.  We define $V'_2=\{(x,t,y) \in V_1 \times V
  \times V_2 \mid \gamma(x,t)=y\}$ and for $q \in Q$, $\gamma'_q:V_1
  \times V \to V'_2$ such that
  $\gamma'_q(x,t)=(x,t,\gamma_q(x,t))$. The dynamics $\gamma'$ behaves
  exactly like $\gamma$ but ``gives'' to the controller the current
  trajectory as this information is stored in the state space $V'_2$.

  We then use $\A_2=(\M, Q,\Goal, \Sigma,\delta',\gamma')$, where
  $\delta'$ is the transition relation $\delta$ adapted to the new
  states $V'_2$: if $(q_1,g,a,R,q_2)\in \delta$ then
  $(q_1,g',a,R',q_2) \in \delta'$ where $g'=\{(x,t,\gamma(x,t)) \mid
  \gamma(x,t) \in g \}$ and for all $(x,t) \in V_1 \times V$,
  $R'(\gamma(x,t))= \{(x',t',\gamma(x',t')) \mid \gamma(x',t') \in
  R(\gamma(x,t))\}$.

  W.l.o.g.  we can suppose that there exists a unique $(x_0,t_0) \in
  V_1 \times V$ such that $\gamma(x_0,t_0)=y$ (if necessary, we add a
  location with constant continuous dynamics pointing to the actual
  location of $y$). Then there exists a
  winning strategy under perfect observation in $\A_1$ from $(q,y)$
  iff there exists a winning strategy under partial observation in
  $\A_2$ from $(q,(x_0,t_0,y))$.  \qed

From the above proposition we get that any definability, decidability,
\textit{etc} result in the partial observation framework will hold in
the perfect observation framework.

\subsection{\texorpdfstring{$\M$}{M}-Games and Bisimulation} \label{subsec:contre_ex}

Time-abstract bisimulation \cite{henzinger95,davoren99,AHLP00} is a
sufficient behavioral relation to check reachability properties of
hybrid systems, and in particular of
$\M$-automata~\cite{brihaye05}. Moreover, it has been shown that it is
also a sufficient behavioral relation in order to solve control
problems in the framework of timed automata~\cite{AMPS98}.  However,
when considering wider classes of hybrid systems, we will see that
this tool is not sufficient anymore for solving control problems in
the perfect observation framework.

\begin{defi}
  Given a mixed transition system $T=(S,\Gamma,\to)$, a {\em
    time-abstract bisimulation for $T$} is an equivalence relation
  $\mathord{\sim} \subseteq S \times S $ such that $\forall q_1,q'_1,
  q_2 \in S$, the two following conditions are satisfied:
  $$\begin{array}{l} \forall a \in \Sigma,\ \left( q_1 \sim q'_1\ 
      \text{and}\ q_1 \xrightarrow{a} q_2 \right) \Rightarrow \\
    \qquad\qquad\qquad \left(
      \exists q'_2 \in S\ \text{s.t.}\ q_2 \sim q'_2\ \text{and}\ q'_1
      \xrightarrow{a} q'_2 \right)\\[0.3cm]
    \forall \tau \in M^+,\ \left( q_1 \sim q'_1\ \text{and}\ q_1
      \xrightarrow{\tau} q_2 \right) \Rightarrow\\
    \qquad \left( \exists \tau' \in M^+,\ \exists
      q'_2 \in S\ \text{s.t.}\ q_2 \sim q'_2\ \text{and}\ q'_1
      \xrightarrow{\tau'} q'_2 \right)
  \end{array}$$
\end{defi}

\begin{exa}\label{bisimnotgood}
  In this example, we assume a perfect observation framework. Let us
  consider the $\M$-game $\A = (\M,Q,\Goal,\Sigma,\delta,\gamma)$
  where $\M = \langle \IR,<,+,0,1,\equiv_2 \rangle$ ($\equiv_2$
  denotes the ``modulo $2$'' relation), $Q=\{q_1,q_2,q_3\}$,
  $\Goal=\{q_2\}$, $\Sigma=\Sigma_c \cup \Sigma_u$ where
  $\Sigma_c=\{c\}$ (resp. $\Sigma_u=\{u\}$) is the set of controllable
  (resp. uncontrollable) actions. The dynamics in $q_1$,
  $\gamma_{q_1}: \IR^+ \times \{0,1\} \times \IR^+ \to \IR^+ \times
  \{0,1\}$ is defined as $\gamma_{q_1}(x_1,x_2,t) = (x_1+t,x_2)$.

  \begin{figure}[ht]

    \null\hfill\hfill\subfigure[The $\M$-game $\A$\label{fig:mgame2}]{
      \begin{tikzpicture}
        \draw (0,0) node [circle,inner sep=1.5pt,draw] (q1) {$q_1$};
        \draw [latex'-] (q1) -- +(-.8,0);
        \draw (2,.8) node [circle,inner sep=1.5pt,draw,fill=black!40!white] (q2) {$q_2$};
        \draw (2,-.8) node [circle,inner sep=1.5pt,draw] (q3) {$q_3$};
        \draw [-latex'] (q1) -- (q2) node [midway,sloped,above] {$g_C,c$};
        \draw [-latex'] (q1) -- (q3) node [midway,sloped,below] {$g_B,u$};
        
        \path (1,-2) --(1,-2);
      \end{tikzpicture}
    }\hfill\hfill
    \subfigure[Dynamics in $q_1$\label{fig:cexbisim}]{
    \begin{tikzpicture}[scale=.8]
      \begin{scope}
        \draw (0,0) -- (5.5,0) node [pos=0,left=.5cm] {$x_2=0$};
        \draw [dotted] (5.5,0) -- (6.5,0);
        \foreach \x in {0,1,2,3,4,5}
        {
          \draw (\x,-.1) -- (\x,.1);
          
          \ifnum\x=1 \path (\x-1,0) -- (\x,0) node [midway,above] {$A$}; \fi
          \ifnum\x=2 \path (\x-1,0) -- (\x,0) node [midway,above] {$C$}; \fi
          \ifnum\x=4 \path (\x-1,0) -- (\x,0) node [midway,above] {$C$}; \fi
          \ifnum\x=3 \path (\x-1,0) -- (\x,0) node [midway,above] {$B$}; \fi
          \ifnum\x=5 \path (\x-1,0) -- (\x,0) node [midway,above] {$B$}; \fi
        }
      \end{scope}
      \begin{scope}[yshift=1cm]
        \draw (0,0) -- (5.5,0) node [pos=0,left=.5cm] {$x_2=1$};
        \draw [dotted] (5.5,0) -- (6.5,0);
        \foreach \x in {0,1,2,3,4,5}
        {
          \draw (\x,-.1) -- (\x,.1);
          
          \ifnum\x=1 \path (\x-1,0) -- (\x,0) node [midway,above] {$A$}; \fi
          \ifnum\x=2 \path (\x-1,0) -- (\x,0) node [midway,above] {$B$}; \fi
          \ifnum\x=4 \path (\x-1,0) -- (\x,0) node [midway,above] {$B$}; \fi
          \ifnum\x=3 \path (\x-1,0) -- (\x,0) node [midway,above] {$C$}; \fi
          \ifnum\x=5 \path (\x-1,0) -- (\x,0) node [midway,above] {$C$}; \fi
        }
      \end{scope}
      \path (0,-2) -- (1,-1);
    \end{tikzpicture} 
  }
  \hfill\hfill\null
    \caption{Time-abstract bisimulation does not preserve winning states}
  \end{figure}
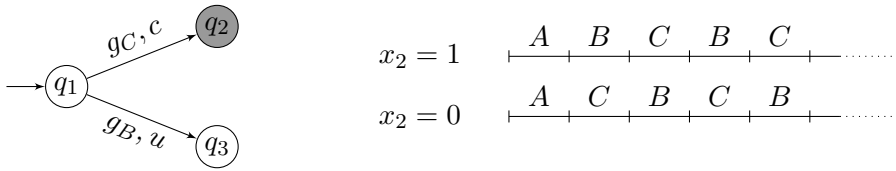
  We consider the partition depicted on Figure \ref{fig:cexbisim}. The
  guard $g_C$ is satisfied on $C$-states and the guard $g_B$ is
  satisfied on $B$-states. Note that this partition is compatible with
  $\Goal$ and w.r.t. discrete transitions.
  
  In this game, the controller can win when it enters a $C$-state by
  performing action $c$ and it loses when entering a $B$-state because
  it cannot prevent the environment from performing a $u$ and going in
  the losing state $q_3$.
  
  It follows that the state $s_1=(q_1,(0,1))$ is losing, whereas the
  state $s_2=(q_1,(0,0))$ is winning. However, the equivalence
  relation induced by the partition $\{A,B,C\}$ is a time-abstract
  bisimulation: the two states $s_1$ and $s_2$ are thus time-abstract
  bisimilar, but not equivalent for the game.  It follows that
  time-abstract bisimulation is not correct for solving control
  problems, in the sense that a time-abstract bisimulation cannot
  always distinguish between winning and losing states.
\end{exa}

\begin{prob}
  \label{prop:incorrect}
  Let $\M$ be a structure and $\A$ an $\M$-game. A partition
  respecting $\Goal$ and inducing a time-abstract bisimulation on $Q
  \times V_2$ does not necessarily respect the set of winning states
  of $\A$.
\end{prob}

\section{The Suffix and the Superword Abstractions}\label{section2}
In this section we explain how to encode symbolically trajectories of
dynamical systems with ``words''. We will present two different
encodings (or abstractions) depending on the observation framework
(perfect or partial) we assume.

\subsection{Perfect Observation and the Suffix  Abstraction}
\label{subsec-suffix}
In this subsection, we review the word encoding technique introduced
in~\cite{BMRT04} in order to study o-minimal hybrid systems. We focus
on the \emph{suffix partition} introduced in \cite{brihaye05}. This
encoding will be suitable in order to study control reachability
problem in the perfect observation framework
(see~Subsection~\ref{subsec-aboutperfect}). We first explain how to
build words associated with trajectories.  Given a dynamical system
$\ds$ and a finite partition $\P$ of $\mkdeux$, given $x \in \mkun$ we
associate a word with the trajectory $\Gamma_x=\{\gamma(x,t) \mid t
\in V\}$ in the following way. We consider the sets $\{t \in V \mid
\gamma(x,t) \in P\}$ for $P \in \P$. This gives a partition of the
time $V$. In order to define a word on $\P$ associated with the
trajectory determined by $x$, we need to define the set of intervals
${\mathcal F}_x = \bigl\{I \mid I\ \text{is a time interval or a point and
  is maximal for the property ``} \exists P \in \P,\ \forall t \in
I,\ \gamma(x,t) \in P\text{''}\bigr\}$.  For each $x$, the set ${\mathcal
  F}_x$ is totally ordered by the order induced from $M$. This allows
us to define \emph{the word on $\P$ associated with the trajectory
  $\Gamma_x$} denoted $\omega_x$.

\begin{defi}\label{def:wordgx}
  Given $x \in \mkun$, \emph{the word associated with $\Gamma_x$} is
  given by the function $\omega_x : {\mathcal F}_x \to \P$ defined by
  $\omega_x(I)= P$, where $I \in {\mathcal F}_x$ is such that $\forall t
  \in I$, $\gamma(x,t) \in P$.
\end{defi}

The set of words associated with $\ds$ over $\P$ gives in some sense a
complete \emph{static} description of the dynamical system $\ds$
through the partition $\P$. In order to recover the \emph{dynamics},
we need further information.

Given a point $x$ of the input space $\mkun$, we have associated with
$x$ a trajectory $\Gamma_x$ and a word $\omega_x$. If we consider
$(x,t)$ a point of the space-time
$\mkun \times V$, it corresponds to a point $\gamma(x,t)$ lying on
$\Gamma_x$. To recover in some sense the position of $\gamma(x,t)$ on
$\Gamma_x$ from $\omega_x$, we associate with $(x,t)$ a suffix of the
word $\omega_x$ denoted $\omega_{(x,t)}$.  The construction of
$\omega_{(x,t)}$ is similar to the construction of $\omega_x$, we only
need to consider the sets of intervals ${\mathcal F}_{(x,t)}= \big\{I \cap
\{t' \in V \mid t' \ge t\} \mid I \in {\mathcal F}_x \big\}.$

Let us notice that given $(x,t)$ a point of the space-time $\mkun
\times V$ there is a unique suffix $\omega_{(x,t)}$ of $\omega_x$
associated with $(x,t)$. Given a point $y \in \mkdeux$ it may have
several $(x,t)$ such that $\gamma(x,t)=y$ and so several suffixes are
associated with $y$. In other words, given $y \in \mkdeux$, the
\emph{future} of $y$ is non-deterministic, and a single suffix
$\omega_{(x,t)}$ is thus not sufficient to recover the dynamics of the
transition system through the partition $\P$. To encode the dynamical
behavior of a point $y$ of the output space $\mkdeux$ through the
partition $\P$, we introduce the notion of suffix abstraction (called
suffix dynamical type in~\cite{brihaye05,brihaye06}) of a point
$y$ w.r.t. $\P$.

\begin{defi}\label{suf}
  Given a dynamical system $\ds$, a finite partition $\P$ of
  $\mkdeux$, a point $y \in \mkdeux$, the \emph{suffix abstraction} of
  $y$ w.r.t. $\P$ is denoted $\Suf_\P(y)$ and defined by $\Suf_\P(y) =
  \{ \omega_{(x,t)} \mid \gamma(x,t) = y\}$.
\end{defi}

This allows us to define an equivalence relation on $\mkdeux$. Given
$y_1$, $y_2 \in \mkdeux$, we say that they are
\emph{suffix-equivalent} if and only if $\Suf_{\P}(y_1) =
\Suf_{\P}(y_2)$.  We denote $\Suf\left(\P\right)$ the partition
induced by this equivalence, which we call the \emph{suffix partition}
w.r.t. $\P$. We say that a partition $\P$ is \emph{suffix-stable} if
$\Suf(\P)=\P$ (it implies that if $y_1$ and $y_2$ belong to the same
piece of $\P$ then $\Suf_{\P}(y_1) = \Suf_{\P}(y_2)$).

To understand the suffix abstraction technique, we provide several
examples.

\begin{exa}
  We start with example~\ref{ex:spiraletot}. The suffix abstraction in
  $(0,0)$ is composed of a unique suffix $ACBC$ because any trajectory
  leaving $(0,0)$ crosses exactly once the spiral at some point. By
  looking at Fig.~\ref{fig:ex-spiraltot} one can convince oneself that
  the suffixes associated with the other points of the plane are given
  by suffixes of $ACBC$; for instance, the points lying on the spiral
  (the piece $B$) have suffix $BC$.
\end{exa}

\begin{exa}\label{ex-ta}
  We first consider a two dimensional timed automata dynamics (see
  Example~\ref{ex:AT}). In this case we have that
  $\gamma(x_1,x_2,t)=(x_1+t,x_2+t)$. We associate with this dynamics
  the partition $\P=\{A,B\}$ where $B=[1,2]^2$ and $A=\IR^2 \setminus
  B$.  In this example the suffix partition is made of three pieces,
  which are depicted in Figure~\ref{figautempo}.
  \begin{figure}[ht]
    \begin{center}
      \begin{tikzpicture}[scale=1.3]
        \draw [latex'-latex'] (0,2.5) -- (0,0) node [pos=0,left] {$x_2$} node [pos=1,below,left] {$0$} -- (2.5,0) node [pos=1,below] {$x_1$};
        \draw [fill=black!20!white] (1,1) -- (2,1) -- (2,2) -- (1,2) --cycle;
        \draw [dashed] (1,0) -- (2,1);
        \draw [dashed] (0,1) -- (1,2);
        \draw (1.5,1.5) node {$BA$};
        \draw (2.5,2.5) node {$A$};
        \draw (.5,.5) node {$ABA$};
      \end{tikzpicture}
    \end{center}
    \caption{Suffixes for the timed automata dynamics}
    \label{figautempo}
  \end{figure}
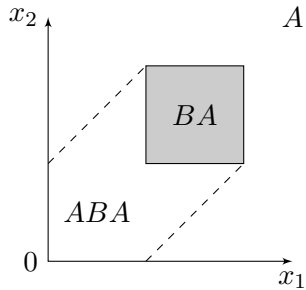
\end{exa}

The suffix abstraction allows to encode more sophisticated continuous
dynamics than the previous suffix encoding of a trajectory.  In the
next example we recover in some sense the continuous dynamics of
\emph{rectangular automata}~\cite{HKPV98}, which requires to use the
suffix abstraction (some of the points do not have a unique suffix).

\begin{exa}\label{ex-rect}
  We consider the dynamical system $({\mathcal M},\gamma)$ where ${\mathcal M}
  = \langle \IR,+,\cdot,0,1,<\rangle$ and $\gamma:\IR^2 \times [1,2]
  \times \IR^+ \to \IR^2$
  is defined by $\gamma(x_1,x_2,p,t)= (x_1 + t, x_2 + p \cdot t)$.  We
  associate with this dynamical system the partition $\P=\{A,B,C\}$
  where $B = [2,5] \times [3,4]$, $C = [3,5]\times[1,2]$ and $A =
  \IR^2 \setminus (B \cup C)$ (see Figure~\ref{fig rect}). Let us
  focus on the suffix abstractions of the two points $y_1 = (1,2.5)$
  and $y_2=(2,0.5)$. We have that $\text{Suf}_{\mathcal P}(y_1) =
  \{A,ABA\}$ and $\text{Suf}_{\mathcal P}(y_2) = \{ABA,ACABA\}$. Though
  several points have several possible suffixes, the partition induced
  by the suffix abstraction is finite and illustrated in
  Figure~\ref{fig rect1}.

  \begin{figure}[ht]
    \null\hfill
    \subfigure[The dynamics\label{fig rect}]{
      \begin{tikzpicture}[scale=.8]
        \draw [latex'-latex'] (0,5.5) -- (0,0) -- (5.5,0);
        \draw [fill=black!60!white,line width=0pt] (3,1) -- (5,1) -- (5,2) -- (3,2) --cycle;
        \draw (4.5,1.5) node {\textcolor{white}{$C$}};
        \draw [fill=black,line width=0pt] (2,3) -- (5,3) -- (5,4) -- (2,4) --cycle;
        \draw (3,3.5) node {\textcolor{white}{$B$}};
        \draw (2.5,5.5) -- (1,2.5) -- (4,5.5);
        \path [opacity=.5,fill=black!20!white,line width=0pt] (2.5,5.5) -- (1,2.5) -- (4,5.5) --cycle;
        \draw (2.5,5.5) -- (1,2.5) -- (4,5.5);
        \draw (1,2.5) [fill=black] circle (1.5pt) node [below left] {$y_1$};
        \path [opacity=.5,fill=black!20!white,line width=0pt] (5.5,5.5) -- (4.5,5.5) -- (2,.5) -- (5.5,4) --cycle;
        \draw (4.5,5.5) -- (2,.5) -- (5.5,4);
        \draw (2,.5) [fill=black] circle (1.5pt) node [below left] {$y_2$};
        \draw (1,1) node {$A$};
      \end{tikzpicture}
    }
    \hfill\hfill
    \subfigure[The suffix partition \label{fig rect1}]{
      \begin{tikzpicture}[scale=.8]
        \path [fill=black!10!white,line width=0pt] (0,0) -- (2,4) -- (0,2) --cycle;
        \path [fill=black!20!white,line width=0pt] (2,0) -- (3,2) -- (1,0) --cycle;
        \draw [latex'-latex'] (0,5.5) -- (0,0) -- (5.5,0);
        \draw (3,1) -- (5,1) -- (5,2) -- (3,2) --cycle;
        \draw (2,3) -- (5,3) -- (5,4) -- (2,4) --cycle;
        \draw (0,2) -- (2,4);
        \draw (0,0) -- (2,4);
        \draw (1,0) -- (3,2);
        \draw (2,0) -- (3,2);
        \draw (2,0) -- (5,3);
        \draw (3.5,0) -- (5,3);
        \draw (4,0) -- (5,1);
        \draw (4.5,0) -- (5,1);
        \draw (1,2.5) [fill=black] circle (1.5pt) node [below left] {$y_1$};
        \draw (2,.5) [fill=black] circle (1.5pt) node [below left] {$y_2$};
        \draw [densely dotted,->] (.5,1.6) .. controls +(-120:20pt) and +(0:20pt) .. (-1,1) node [pos=1,left] {${\scriptstyle \{A,ABA\}}$};
        \draw [densely dotted,->] (2,.8) .. controls +(90:20pt) and +(-120:20pt) .. (2.7,2.2) node [pos=1,above] {${\scriptstyle \{ABA,ACABA\}}$};
      \end{tikzpicture}
    } \hfill\null
    \caption{A rectangular dynamics}
  \end{figure}
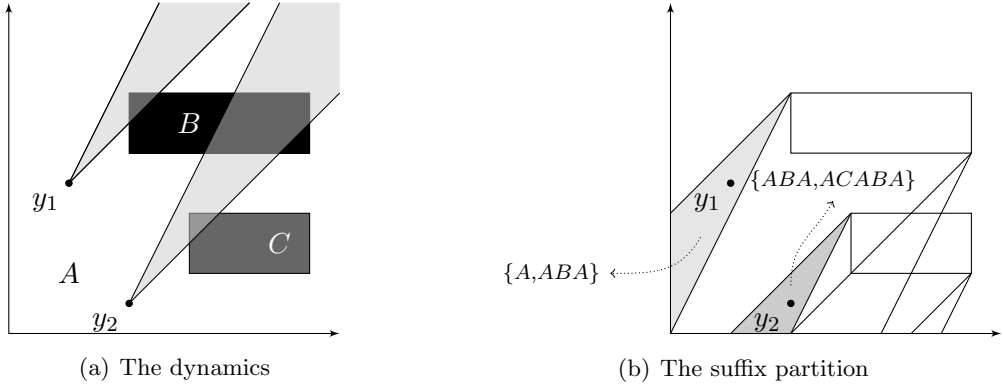

\end{exa}

\subsection{Partial Observation and the Superword Abstraction}

The suffix-partition proposed in Subsection~\ref{subsec-suffix} is not
suitable for the partial observation framework. We will intuitively
convince the reader of this fact.  Let $({\mathcal M},\gamma)$ be a
dynamical system, $y$ be a point of $V_2$ and $\P$ be a partition of
$V_2$. Since several trajectories cross the point $y$, there exist
several $y'$ such that $y \xrightarrow{\tau} y'$, for some $\tau \in
M^+$. In the partial observation framework, the controller does not
know which trajectory will be chosen by the environment and have to
choose a pair $(\tau,c)$ independently. In particular, starting from
$y$, one can potentially be in several different pieces of $\P$ after
$\tau$ time units. The notion of suffix abstraction is not sufficient
in order to capture these behaviors, that is why we now associate a
word $\omega_y$ on $2^{\mathcal P}$ with a given $y \in V_2$. We will see
in~Subsection~\ref{subsec-partial2} that this new encoding is suitable
in order to study control reachability problem in the partial
observation framework.  In order to define the word on $2^{\mathcal P}$
associated with $y \in V_2$, we need to introduce further definitions.

\begin{defi}
  Let $y$ be a point of $V_2$ and $\tau$ be a time in $M^+$.
  \[
  {\mathcal F}_y(\tau) = \big\{ P \in {\mathcal P} \mid \exists x \in M^{k_1}
  \ \exists t \in M \ \gamma(x,t)=y \text{ and } \gamma(x,t+\tau) \in
  P \big\} .
  \]
  The set ${\mathcal F}_y(\tau)$ represents the set of pieces that we have
  potentially reached after $\tau$ time units when starting from $y$.
\end{defi}

\begin{defi}
  Let $y$ be a point of $V_2$.
  \begin{align*}
    {\mathcal F}_y = & \big\{ I\ \mid\ I \text{ is a time interval and
      is maximal for the property }\\
    & \qquad \qquad \qquad \exists S \in 2^{\mathcal P} \ \forall \tau \in
    I \ \ {\mathcal F}_y(\tau) = S \big\}
  \end{align*}
\end{defi}
For each $y \in V_2$, the set ${\mathcal F}_y$ exactly consists of the
connected components of the sets $\{ \tau \in M^+ \mid {\mathcal
  F}_y(\tau) = S \}$, for $S \in 2^{\mathcal P}$.  We can now define the
superword $\Sup_{\P}(y)$ associated with a given $y \in V_2$.

\begin{defi}
  Let $({\mathcal M},\gamma)$ be a dynamical system, $y$ be a point of
  $V_2$, and $\mathcal P$ be a partition of $V_2$. \emph{The superword
    associated with $y$} is given by the function $\Sup_{\P}(y):{\mathcal
    F}_y \to 2^{\mathcal P}$ defined by:
  \[
  \Sup_{\P}(y)(I) = S \qquad \text{ where } I \in {\mathcal F}_y \text{ is
    such that } \forall \tau \in I \ \ {\mathcal F}_y(\tau) = S.
  \]
\end{defi}

Let us notice that given $(\M,\gamma)$ a dynamical system, $\P$ a
partition of $V_2$, and $y$ a point of $V_2$, there exists a unique
superword $\Sup_{\P}(y)$ associated with $y$.  If $(\M,\gamma)$ is a
dynamical system and $\P$ a finite partition of $V_2$, we write
$\Sup(\P)$ for the partition induced by superwords.  We say that a
partition $\P$ is \emph{superword-stable} if $\Sup(\P)=\P$. Let us
illustrate this new notion on examples.

\begin{exa}\label{ex:supword}
  Let us consider the three dynamical systems depicted on
  Figures~\ref{sufsup}. In the three cases, the dynamical system
  consists of two trajectories exiting the point $y_i$. What differs
  in the three systems is the way the partition $\P=\{A,B,C\}$ is
  crossed. We are interested in the superword associated with
  $y_i$. For the two first dynamical systems we have that
  $\Sup_{\P}(y_1) = \Sup_{\P}(y_2) = \{A\}\{B,C\}$, and for the last
  one we have that $\Sup_{\P}(y_3)=\{A\}\{B,C\} \{B\} \{B,C\} \{C\}
  \{B,C\}$.

  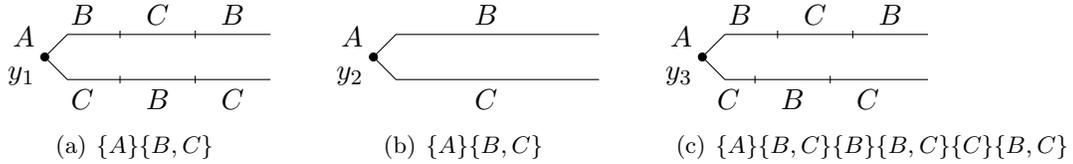
\begin{figure}[h]
    \null\hfill\subfigure[$\{A\}\{B,C\}$]{
      \begin{tikzpicture}
        \draw (0,0) [fill=black] circle (1.5pt) node [below left] {$y_1$} node [above left] {$A$};
        \draw (3,-.3) -- (.3,-.3) -- (0,0) -- (.3,.3) -- (3,.3);
        \path (.3,.3) -- (.7,.3) node [midway,above] {$B$};
        \path (1,.3) -- (2,.3) node [midway,above] {$C$};
        \path (2,.3) -- (3,.3) node [midway,above] {$B$};
        \draw (1,.25) -- (1,.35);
        \draw (2,.25) -- (2,.35);
        \path (.3,-.3) -- (.7,-.3) node [midway,below] {$C$};
        \path (1,-.3) -- (2,-.3) node [midway,below] {$B$};
        \path (2,-.3) -- (3,-.3) node [midway,below] {$C$};
        \draw (1,-.25) -- (1,-.35);
        \draw (2,-.25) -- (2,-.35);
      \end{tikzpicture}
}\hfill
    \subfigure[$\{A\}\{B,C\}$]{
      \begin{tikzpicture}
        \draw (0,0) [fill=black] circle (1.5pt) node [below left] {$y_2$} node [above left] {$A$};
        \draw (3,-.3) -- (.3,-.3) -- (0,0) -- (.3,.3) -- (3,.3);
        \path (.3,.3) -- (2.7,.3) node [midway,above] {$B$};
        \path (.3,-.3) -- (2.7,-.3) node [midway,below] {$C$};
      \end{tikzpicture}
}\hfill
    \subfigure[\mbox{$\{A\}\{B,C\} \{B\} \{B,C\} \{C\} \{B,C\}$}]{
      \begin{tikzpicture}
        \draw (0,0) [fill=black] circle (1.5pt) node [below left] {$y_3$} node [above left] {$A$};
        \draw (3,-.3) -- (.3,-.3) -- (0,0) -- (.3,.3) -- (3,.3);
        \path (.3,.3) -- (.7,.3) node [midway,above] {$B$};
        \path (1,.3) -- (2,.3) node [midway,above] {$C$};
        \path (2,.3) -- (3,.3) node [midway,above] {$B$};
        \draw (1,.25) -- (1,.35);
        \draw (2,.25) -- (2,.35);
        \path (.3,-.3) -- (.4,-.3) node [midway,below] {$C$};
        \path (.7,-.3) -- (1.7,-.3) node [midway,below] {$B$};
        \path (1.7,-.3) -- (2.7,-.3) node [midway,below] {$C$};
        \draw (.7,-.25) -- (.7,-.35);
        \draw (1.7,-.25) -- (1.7,-.35);
      \end{tikzpicture} \hspace*{2cm}
      }
    \caption{Suffix and superword are not comparable}
    \label{sufsup}
  \end{figure}

  Let us notice that the notions of \emph{suffix abstraction} and
  \emph{superword abstraction} are incomparable. To illustrate this
  fact, let us consider again the three dynamical systems of
  Figure~\ref{sufsup}. We have that
  $\Sup_{\P}(y_1)=\Sup_{\P}(y_2)\ne\Sup_{\P}(y_3)$. Let us now
  consider the suffix abstractions of these points:
  \[
  \Suf(y_1) = \{ ABCB,ACBC\} \ ; \ \Suf(y_2) = \{ AB,AC\} \ ; \
  \Suf(y_3) = \{ ABCB,ACBC\}.
  \]
  This shows that the superword abstraction can distinguish between
  $y_1$ and $y_3$, but cannot distinguish between $y_1$ and $y_2$,
  although the suffix abstraction can distinguish between $y_1$ and
  $y_2$, but cannot distinguish between $y_1$ and $y_3$.
\end{exa} 

\section{Solving an \texorpdfstring{$\M$}{M}-Game}\label{mgame}

In this section we first present a general procedure to compute the
set of winning states for an $\M$-game under partial observation. We
then show that if a partition is superword-stable,
the procedure can be performed symbolically on pieces of the
partition. The procedure described is not always effective and we will
later point out specific $\M$-structures for which each step of the
procedure is computable.  By Proposition~\ref{prop:casparticulier}, we
know that the perfect observation control problem can be seen as a
special case of the partial observation framework; however at the end
of this section, we explain how the suffix partition can be used in
order to directly solve the perfect observation control problem.

\subsection{Controllable Predecessors under Partial Observation}
\label{subsec:contpred}

As for classical reachability games~\cite{lncs2500}, one way of
computing winning states is to compute the {\em attractor} of goal
states by iterating a {\em controllable predecessor} operator.  Let
$\A=(\M,Q,\Goal,\Sigma,\delta,\gamma)$ be an $\M$-game. For $W
\subseteq Q \times V_2$, 
$a \in \Sigma_c$ and $u \in \Sigma_u$ we first define the notion of
controllable discrete predecessors. For every $a \in \Sigma = \Sigma_c
\cup \Sigma_u$, we have
$$\Pred_a(W) = \left\{(q,y) \in Q \times V_2 \ \begin{array}{c}
    \mid\\[-0.2cm] \mid \\[-0.2cm] \mid\\[-0.2cm] \mid \\[-0.2cm] \mid
  \end{array}\ 
  \begin{array}{l} a\ \text{is enabled in}\ (q,y),  \\
    \text{and}\ \forall (q',y') \in Q \times V_2,\ \\
    \left((q,y) \xrightarrow{a} (q',y') \Rightarrow (q',y') \in
      W\right) \end{array}\right\}.$$ The
intuition of this operator is the following: a state is in
$\Pred_a(W)$ if action $a$ can be done from $(q,y)$, and whichever
transition is taken leads to a state in $W$ (action $a$ ensures $W$ in
one step).  We also define $\cPred(W) = \displaystyle \bigcup_{c \in
  \Sigma_c} \Pred_c(W)$ and $\uPred(W) = \displaystyle \bigcup_{u \in
  \Sigma_u} \Pred_u(W)$.

As for timed and hybrid games~\cite{AMPS98,HHM99}, we also define a
\emph{safe time predecessor} of a set $W$ w.r.t. a set $W'$, that is
specific to the partial observation framework:
a state $(q,y)$ is in $\timePred_{\textsf{partial}}(W,W')$ if a delay
$\tau$ can be chosen such that for all trajectories starting from
$(q,y)$, one can let $\tau$ time units pass avoiding $W'$ and then
reach $(q',y') \in W$. Formally the operator
$\timePred_{\textsf{partial}}$ is defined as follows:
$$\timePred_{\textsf{partial}}(W,W') = \left\{ (q,y) \in Q \times V_2
\begin{array}{c}
    \mid\\[-0.2cm] \mid \\[-0.2cm] \mid\\[-0.2cm] \mid \\[-0.2cm] \mid
  \end{array}\!\!
  \begin{array}{l} \!\!\exists \tau \in M^+,\ \forall (x,t) \in V_1 \times
    V\ \text{s.t.}\\ \!\!\gamma_q(x,t)=y,\ \text{and}\ (q,y)
    \xrightarrow{\tau}_{x,t} (q',y')\\ \!\!\text{implies} \left((q',y')\in W\
    \text{and}\ \Post_{[t,t+\tau]}^{q,x} \subseteq
    \overline{W'}\right)\!\!\end{array}\right\}.$$ where
$\Post_{[t,t+\tau]}^{q,x} = \{\gamma_q(x,t')\ \mid\ t \le t' \le
t+\tau \}$.

\medskip The \emph{controllable predecessor} operator under partial
observation $\pi_{\textsf{partial}}$ is then defined as:
\[
\pi_{\textsf{partial}}(W)= W \cup \bigcup\limits_{a \in \Sigma_c}
\timePred_{\textsf{partial}}(\Pred_a(W), \uPred(\overline{W})).
\]

\begin{rem}
  Note that the operator $\pi_{\textsf{partial}}$ is definable in any
  expansion of an ordered group. Hence, if $W$ is definable, so is
  $\pi_{\textsf{partial}}(W)$.
\end{rem}

\begin{exa}
  We first illustrate the computation of the operator
  $\pi_{\textsf{partial}}$ on Example~\ref{ex:spiraletot} (see
  page~\pageref{ex:spiraletot}). In this case,
  $\pi_{\textsf{partial}}$ does not induce a winning strategy from
  $(q_1,(0,0))$ under partial observation. Setting $W = \Goal \times
  V_2 = \{q_2\} \times V_2$, we have that $\pi_{\textsf{partial}}(W)$
  does not contain the point $(q_1,(0,0))$ because there is no uniform
  choice for a positive delay $\tau$ before taking action $c$ so that
  the spiral (area $B$) can be avoided. Notice however that
  $\pi_{\textsf{partial}}(W)$ is not empty because it includes all
  points different from $(q_1,(0,0))$ (from which there is a unique
  trajectory). 
\end{exa}

\begin{rem}
  \label{rk:partial}
  Note also that due to the partial observation assumption, in the
  definition of $\pi_{\textsf{partial}}$, the action $a$ for
  controlling the system has to be chosen before choosing the delay
  $\tau$. Indeed, the controller does not know which precise
  trajectory will be chosen by the environment, in particular, action
  $a$ should be available after time $\tau$ independently of the
  choice of trajectory made by the environment. This is illustrated in
  the next example.
\end{rem}

\begin{exa}\label{ex:ab}
  Let us consider the $\M$-game $\A$ depicted on
  Figure~\ref{fig:mgamebis} where $\Goal=\{q_2,q_3\}$ and where $c_1,
  c_2 \in \Sigma_c$ are distinct controllable actions. The dynamics in
  $q_1$ is depicted on Figure~\ref{fig:dynqun}, roughly speaking, it
  consists of of two trajectories exiting the point $y$.  perfect
  observation from $y$; indeed depending on the trajectory we are
  following, we will either play $(\tau,c_1)$ or $(\tau,c_2)$, for
  some well-chosen $\tau \in \IR^+$. However, there is no winning
  strategy under partial observation from $y$. Although we can find
  $\tau \in \IR^+$ such that a controllable action will be (safely)
  available (from $y$) after $\tau$ time units, we are unable to tell
  which controllable action will be taken.

  In fact if  $W=\Goal \times V_2$ we have that
  $\pi_{\textsf{partial}}(W)= \{(q_1,z) \mid z \in V_2\backslash \{y\}\}$.
  Indeed if $(q_1,z) \neq (q_1,y)$, the controller can deduce the trajectory
  from the current state and choose its action accordingly.

  \begin{figure}[ht]
    \null\hfill\subfigure[The $\M$-game $\A$\label{fig:mgamebis}]{
      \begin{tikzpicture}
        \draw (0,0) node [circle,inner sep=1.5pt,draw] (q1) {$q_1$};
        \draw [latex'-] (q1) -- +(-.8,0);
        \draw (2,.8) node [circle,inner sep=1.5pt,draw,fill=black!40!white] (q2) {$q_2$};
        \draw (2,-.8) node [circle,inner sep=1.5pt,draw,fill=black!40!white] (q3) {$q_3$};
        \draw [-latex'] (q1) -- (q2) node [midway,sloped,above] {$g_B,c_1$};
        \draw [-latex'] (q1) -- (q3) node [midway,sloped,below] {$g_C,c_2$};
      \end{tikzpicture}
    }\hfill\hfill
    \subfigure[Dynamics in $q_1$\label{fig:dynqun}]{
      \begin{tikzpicture}
        \draw (0,0) [fill=black] circle (1.5pt) node [below left] {$y$} node [above left] {$A$};
        \draw (3,-.3) -- (.3,-.3) -- (0,0) -- (.3,.3) -- (3,.3);
        \path (1.5,.3) -- (3,.3) node [midway,above] {$B$};
        \draw (1.5,.25) -- (1.5,.35);
        \path (1.5,-.3) -- (3,-.3) node [midway,below] {$C$};
        \draw (1.5,-.25) -- (1.5,-.35);
        \path (0,-.8) -- (1,-.8);
      \end{tikzpicture}}\hfill\null
    \caption{}
  \end{figure}
\end{exa}

The next proposition states the soundness of this operator for
computing winning states in the games under a partial observation
hypothesis.

\begin{prob}\label{proppi2}
  Let $\A=(\M,Q,\Goal,\Sigma,\delta,\gamma)$ be an $\M$-game. If there
  exists $n\in \IN$ s.t. $\pi_{\textsf{\upshape
      partial}}^n(\Goal)=\pi_{\textsf{\upshape partial}}^{n+1}(\Goal)$
  then $\pi_{\textsf{\upshape partial}}^*(\Goal)=\pi_{\textsf{\upshape
      partial}}^n(\Goal)$ is the set of winning states of~$\A$ under
  partial observation.
\end{prob}

\proof
  We first prove that if $(q,y) \in \pi_{\textsf{\upshape
      partial}}^*(\Goal)$ then there exists a winning strategy under
  partial observation from $(q,y)$. To this aim, we define a
  memoryless winning strategy from any $(q,y) \in
  \pi_{\textsf{\upshape partial}}^*(\Goal)$. By notation misuse, we
  define the strategy $\lambda$ on states $(q,y)$ instead of
  executions.
  
  We define a strategy $\lambda$ on all sets $\bigcup_{0 \le i \le k}
  \pi_{\textsf{\upshape partial}}^i(\Goal)$ by induction on $k$, and
  prove that it is a winning strategy. If $k=0$, we assume $\lambda$
  is defined nowhere, it is thus winning from all states in
  $\Goal$. 

  Suppose now that $\lambda$ is already defined on $W=\bigcup_{0 \le i
    \le k} \pi_{\textsf{\upshape partial}}^i(\Goal)$ and is winning on
  these states. We now define $\lambda$ on $\pi_{\textsf{\upshape
      partial}}(W)$. Let $(q,y) \in Q \times V_2$: if $(q,y) \in W$,
  $\lambda$ is already defined; if $(q,y) \in \pi_{\textsf{\upshape
      partial}}(W) \setminus W$, then we know that there exists $a\in
  \Sigma_c$ with $(q,y) \in
  \timePred_{\textsf{partial}}\left(\Pred_a(W), \uPred(\overline{W})
  \right)$. There exists $\tau \in M^+$ with $(\tau,a)$
  enabled\footnote{We say that $(\tau,a) \in M^+ \times \Sigma$ is
    enabled in $(q,y)$ if there exists $(x,t) \in V_1 \times V$ such
    that $\gamma(x,t)=y$ and $(\tau,a)$ is enabled in $(q,x,t,y)$.} in
  $(q,y)$ such that for every $(x,t)$ if $\gamma_q(x,t)=y$, then
  $(q,y) \xrightarrow{\tau,a}_{x,t} (q',y')$, $(q',y') \in W$ and
  $\Post_{[t,t+\tau]}^{q,x} \subseteq
  \overline{\uPred{(\overline{W}})}$.  We set $\lambda(q,y)=(\tau,a)$
  and show that this is a winning choice.
 
  We show by induction on $k$ that $\lambda$ is winning for each
  state of $W=\bigcup_{0 \le i \le k} \pi_{\textsf{\upshape
      partial}}^i(\Goal)$.  This is immediate for $k=0$. Suppose now
  that the result is true for $k$ and let $(q,y) \in
  \pi_{\textsf{\upshape partial}}(W)$. Let $\rho = (q,x,t,y)
  \xrightarrow{\tau_1,a_1} (q_1,x_1,t_1,y_1) \xrightarrow{\tau_2,a_2}
  \ldots$ be an execution compatible with $\lambda$. We have that
  either $\tau_1 = \tau$ and $a_1=a$, in which case $(q_1,y_1)\in W$,
  or $\tau_1 \le \tau$ and $a_1 \in \Sigma_u$, in which case $(q,y)
  \xrightarrow{\tau_1}_{x,t} (q',y') \xrightarrow{a_1} (q_1, y_1)$
  with $(q',y') \notin \uPred{(\overline{W})}$ so $(q_1,y_1) \in
  W$. In both cases, $(q_1,y_1) \in W$ so by induction hypothesis,
  $\rho$ is winning.

  \medskip We now show that if there exists a strategy under partial
  observation $\lambda$ winning from $(q,y)$ then $(q,y) \in
  \pi_{\textsf{\upshape partial}}^*(\Goal)$. Set
  $W=\pi_{\textsf{\upshape partial}}^*(\Goal)$, by contradiction
  suppose that $(q,y) \notin W$, we will construct a non-winning
  execution compatible with $\lambda$. By hypothesis
  $\pi_{\textsf{\upshape partial}}(W)=W$ so $(q,y) \notin
  \pi_{\textsf{\upshape partial}}(W)$, it follows that for all $a \in
  \Sigma_c$, for all $\tau \in M^+$ there exists $(x,t) \in V_1 \times
  V$ such that $\gamma_q(x,t)=y$, and $(q,y) \to_{x,t}^{\tau} (q',y')$
  implies $(q',y')\notin \Pred_a(W)$ or $\Post_{[t,t+\tau]}^{q,x} \cap
  \uPred(\overline{W}) \ne \emptyset$. Let $(\tau,a) = \lambda(q,y)$
  (as $\lambda$ is a strategy under partial observation it does not
  depend of $x$ and $t$) and let $(x,t) \in V_1 \times M^+$ be as in
  the previous statement.

  There exists $(q_1,x_1,t_1,y_1)$ with $(q_1,y_1) \notin W$ such that
  either $(q,x,t,y) \xrightarrow{\!\tau,a\!} (q_1,x_1,t_1,y_1)$ or there
  exists $\tau' \le \tau$ and $u \in \Sigma_u$ with $(q,x,t,y)
  \xrightarrow{\tau',u} (q_1,x_1,t_1,y_1)$. In both cases, the
  constructed execution is compatible with $\lambda$. As $(q_1,y_1)
  \notin W$ we can repeat the same argument and construct inductively
  an execution $\rho =(q,x,t,y) \xrightarrow{\tau_1,a_1}
  (q_1,x_1,t_1,y_1) \xrightarrow{\tau_2,a_2} \ldots$ compatible with
  $\lambda$ and such that for every $i$, $(q_i,x_i,t_i,y_i) \notin W$.
  By definition of $W$, for every $i$, $q_i \notin \Goal$, which
  contradicts the assumption that $\lambda$ is a winning strategy.
  \qed

$\pi_{\textsf{\upshape partial}}^*(\Goal)$, but this does not imply
that we can compute this set, as some $\M$-structures have an
undecidable theory.  The following corollary states that if some
conditions on the structure and on $\pi_{\textsf{\upshape partial}}$
are satisfied, then this procedure provides an algorithmic solution to
the control problem.

\begin{cor} \label{cor:etatsgagnants} Let $\M$ be a structure
  such that $\Th(\M)$ is decidable.\footnote{We recall that a theory
    $\Th(\M)$ is decidable iff there is an algorithm which can
    determine whether or not any sentence ({\it i.e.}, a formula with
    no free variable.) is a member of the theory ({\it i.e.}, is
    true). We suggest to readers interested in general decidability
    issues on o-minimal hybrid systems to refer to Section~5
    of~\cite{BM05}.}  Let $\C$ be a class of $\M$-games such that for
  every $\A$ in $\C$, there exists a finite partition $\P$ of $Q
  \times \mkdeux$ definable in $\M$, respecting
  $\Goal$\footnote{\emph{I.e.}, $\Goal$ is a union of pieces of
    $\P$.}, and stable under $\pi_{\textsf{\upshape
      partial}}$.\footnote{Meaning that if $P$ is a piece of $\P$ then
    $\pi_{\textsf{\upshape partial}}(P)$ is a union of pieces of
    $\P$.}  Then the control problem under partial observation in the
  class $\C$ is decidable.  Moreover if $\A \in \C$, the set of
  winning states under partial observation of $\A$ is computable.
\end{cor}

\proof
  Let $\M$ be a structure and $\C$ a class of automata satisfying the
  hypotheses and take $\A \in \C$.  As $\P$ is stable under
  $\pi_{\textsf{\upshape partial}}$, $\pi_{\textsf{\upshape
      partial}}^*(\Goal)$ is a finite union of pieces of $\P$.  Hence
  there exists $n \in \IN$ such that $\pi_{\textsf{\upshape
      partial}}^*(\Goal) = \pi_{\textsf{\upshape
      partial}}^n(\Goal)$. Thus proposition \ref{proppi2} shows that
  the set of winning states is $\pi_{\textsf{\upshape
      partial}}^*(\Goal)$.

  As $\pi_{\textsf{\upshape partial}}$ and $\Goal$ are definable, we
  have that $\pi_{\textsf{\upshape partial}}^i(\Goal)$ is definable
  and as $\Th(\M)$ is decidable we can test if $\pi_{\textsf{\upshape
      partial}}^i(\Goal)=\pi_{\textsf{\upshape
      partial}}^{i+1}(\Goal)$, we can thus effectively find a
  representation of $\pi_{\textsf{\upshape partial}}^*(\Goal)$.
  
  As $\Th(\M)$ is decidable, if a state $(q,y)$ is definable we can
  test if $(q,y) \in \pi_{\textsf{\upshape partial}}^*(\Goal)$. It
  follows that the control problem in an $\M$-structure is decidable.
  \qed

\subsection{Superwords and the \texorpdfstring{$\pi_{\textsf{\upshape partial}}$}{pi_Partial} Operator}\label{subsec-partial2}
We now present a sufficient condition for a partition to be stable
under the operator $\pi_{\textsf{\upshape partial}}$: we require that
the partition is stable under $\Pred_a$ (for all $a \in \Sigma$) to
handle the discrete part of the automaton and we show that the
stability by superwords is fine enough to be correct for solving
control problems under partial observation.

\begin{prob} \label{prop:supwordstable} Let $\A$ be an
  $\M$-game and $\P$ be a partition of $Q \times V_2$.  If $\P$
  respects $\Goal$, is stable under $\Pred_a$ (for all $a \in \Sigma$)
  and superword-stable, then $\P$ is stable under the operator
  $\pi_{\textsf{\upshape partial}}$.
\end{prob}

\proof 
  We fix a location $q$ of the automaton and we take $y_1,y_2 \in V_2$
  such that there exists $A \in \P$ with $y_1,y_2 \in A$. We now show
  that if $y_1 \in \pi_{\textsf{\upshape partial}}(X)$, for some $X
  \in \P$ then $y_2 \in \pi_{\textsf{\upshape partial}}(X)$. In case
  $y_1 \in X$ then $X = A$ and $y_2 \in Y$.

  We assume $y_1 \in \pi_{\textsf{partial}}(X) \setminus X$.  
  There exists $a \in \Sigma_c$ and $\tau_1 \in M^+$ such that for all
  $(x,t) \in V_1 \times V$ with $\gamma_q(x,t)=y_1$ and for all $y'_1$
  such that $y_1 \xrightarrow{\tau_1}_{x,t} y'_1$, we have that $y'_1
  \in \Pred_a(X)$, and $\Post_{[t,t+\tau_1]}^{q,x} \subseteq
  \overline{\uPred{(\overline{X}})}$. Let us now express the previous
  condition in term of superword. Assume that
  $$\Sup_{\P}(y_1) = S_1S_2\cdots S_k, \quad 
  \text{ where } S_i \in 2^{\P},$$ the previous condition means that
  $\Sup_{\P}(y_1)$ contains a prefix $S_1 \cdots S_l$ is such that:
  \begin{itemize}
  \item for all $P_i \in S_l$, we have that $P_i \subseteq \Pred_a(X)$
    (this condition makes sense since $\mathcal P$ is stable under
    $\Pred_a$; indeed, \textit{a priori} we only have that there
    exists $y'_1 \in P_i$ such that $y'_1 \in \Pred_a(X)$, the
    stability of $\mathcal P$ under $\Pred_a$ implies that $P_i \subseteq
    \Pred_a(X)$),
  \item for all $j \le l$, for all $P_i \in S_j$, we have that
    $\uPred(\overline{X}) \cap P_i = \emptyset$ (again this condition
    makes sense since $\mathcal P$ is stable under $\Pred_a$).
  \end{itemize}
  
  Since $\P= \Sup\left(\P\right)$ and both $y_1$ and $y_2$ belong to
  the same piece of $\P$, we have that $\Sup_\P(y_1)=\Sup_\P(y_2)=
  S_1S_2\cdots S_k$.  In particular, we can find $\tau_2 \in M^+$ such
  that if $y_2 \xrightarrow{\tau_2} y'_2$, we have that $y'_2$
  corresponds to the letter $S_l$.  Thus we have that $y'_2 \in
  \Pred_a(X)$ and $\Post_{[t,t+\tau_2]}^{q,x} \subseteq
  \overline{\uPred{(\overline{X}})}$, \textit{i.e.} $y_2 \in
  \pi_{\textsf{\upshape partial}}(X)$.  \qed

As an immediate corollary of this proposition and of
Corollary~\ref{cor:etatsgagnants}, we get the following general
decidability result.

\begin{cor} \label{corta} Let $\M$ be a structure such that
  $\Th(\M)$ is decidable. Let $\C$ be a class of $\M$-games such that
  for every $\A$ in $\C$, there exists a finite partition $\P$ of $Q
  \times \mkdeux$ definable in $\M$, respecting $\Goal$,
  superword-stable, and
  stable under $\Pred_a$ for every action $a \in \Sigma$. Then the
  control problem under partial observation
  (Problem~\ref{prob:contpartial}) in the class $\C$ is decidable, and
  if $\A \in \C$, the set of winning states under partial observation
  of $\A$ is computable.
\end{cor}

\subsection{A Note on the Perfect Observation
  Framework}\label{subsec-aboutperfect} We briefly discuss the perfect
observation framework. We have already seen that it is a special case
of the partial observation framework (see
Proposition~\ref{prop:casparticulier}). Hence, we can reuse the
previous results and get decidability and computability results.
However, we can also define an appropriate controllable predecessor
operator $\pi_{\textsf{perfect}}$ that will be correct in the perfect
observation framework. The new operator $\pi_{\textsf{perfect}}$ is
just a twist of the previous operator, which we define as:
\[
\pi_{\textsf{perfect}}(W) = W \cup
\timePred_{\textsf{perfect}}\left(\cPred(W),\uPred(\overline{W})\right)
\]
where $\timePred_{\textsf{perfect}}$ existentially quantifies on pairs
$(x,t)$ such that $y = \gamma_q(x,t)$ (instead of universally
quantifying on those pairs, as in $\timePred_{\textsf{partial}}$).
\begin{rem}
  In the perfect observation framework, the controller is aware of the
  precise trajectory that will be followed, hence his choice of action
  can be done after his choice of delay contrarily to the partial
  observation case (remember Remark~\ref{rk:partial}).  That is why
  the union over actions is put within the scope of the safe time
  predecessor in $\pi_{\textsf{perfect}}$.
\end{rem}

Applying similar reasoning as in the previous sections, we can prove
that $\pi_{\textsf{perfect}}^*(\Goal)$ corresponds to the set of
winning states of $\A$, and that a partition, which is both stable
under $\Pred_a$ (for every $a \in \Sigma$) and suffix-stable, is
actually correct for solving control problems in the perfect
observation framework. We can thus state the following theorem.

\begin{thm} \label{thmta} Let $\M$ be a structure such that
  $\Th(\M)$ is decidable. Let $\C$ be a class of $\M$-games such that
  for every $\A$ in $\C$, there exists a finite partition $\P$ of $Q
  \times \mkdeux$ definable in $\M$, respecting $\Goal$,
  suffix-stable, and stable under $\Pred_a$ for every action $a \in
  \Sigma$. Then the control problem under perfect observation
  (Problem~\ref{prob:controlperfect}) in the class $\C$ is decidable,
  and if $\A \in \C$, the set of winning states under perfect
  observation of $\A$ is computable.
\end{thm}

Note that being suffix-stable is a stronger condition than being a
time-abstract bisimulation~\cite{brihaye05}, and we see here that this
is one of the right tools to solve control problems. For instance in
Example~\ref{bisimnotgood} the partition $\mathcal P$ is a time-abstract
bisimulation but is not suffix-stable. Indeed $s_1,s_2 \in A$ but
$\text{Suf}_{\mathcal P}(s_1) \ne \text{Suf}_{\mathcal P}(s_2)$.

\begin{rem}\label{remark:TA}
  Using the results of this section, we recover the results of
  \cite{AMPS98} about control of timed automata. Note that for the
  timed automata dynamics (remember Example~\ref{ex:AT}) partial or
  perfect observation do not make a difference (the dynamics is
  deterministic).  Indeed we consider the classical finite partition
  of timed automata that induces the region graph
  (see~\cite{AD94}). Let us call ${\mathcal P}_R$ this partition, and
  notice that ${\mathcal P}_R$ is definable in $\langle
  \IR,<,+,0,1\rangle$.  ${\mathcal P}_R$ is stable under the action of
  $\Pred_a$ for every action $a \in \Sigma$. By Example~\ref{ex:AT}
  the continuous dynamics of timed automata is definable in $\langle
  \IR,<,+,0,1\rangle$. Hence it makes sense to encode continuous
  trajectories of timed automata as words.  Then one can easily verify
  that $\Suf(\P_R)=\P_R$. By Theorem~\ref{thmta} we get the
  decidability and computability of winning states under perfect
  information in timed games \cite{AMPS98} as a side result.
\end{rem}

\begin{cor}
  The control problem under perfect information in the class of timed
  automata is decidable. Moreover the set of winning states under
  perfect observation is computable.
\end{cor}

\section{O-Minimal Games}\label{omingames}

In this section, we focus on the particular case of o-minimal games
({\it i.e.}, $\M$-games where $\M$ is an o-minimal structure and in
which extra assumptions are made on the resets). We first briefly
recall definitions and results related to o-minimality~\cite{PS86}.
We show that existence of finite partitions which are stable
w.r.t. the controllable predecessor operator can be guaranteed for
o-minimal games. More precisely, we first show that, in this
framework, a partition stable under the controllable predecessor
operator can easily be obtained \textit{via} the superword abstraction
(this is due to the assumptions on the resets). Then, we use
properties of o-minimality to prove the finiteness of the previously
obtained partition.  Finally we focus on o-minimal structures with a
decidable theory in order to obtain full decidability and
computability results. As in the previous section, we mostly focus on
the partial observation framework, but also mention results in the
perfect observation framework.

\subsection{O-Minimality}
We recall here the definition of o-minimality and the ``\emph{Uniform
  Finiteness Theorem}'' that will be applied later in this
section. The reader interested in o-minimality should refer
to~\cite{vandendries98} for further results and an extensive
bibliography on this subject.

\begin{defi} \label{o-minimal} An extension of an ordered
  structure ${\mathcal M} = \langle M, \mathord< , \ldots \rangle$ is
  \emph{o-minimal} if every definable subset of $M$ is a finite union
  of points and open intervals (possibly unbounded).
\end{defi}

In other words the definable subsets of $M$ are the simplest possible:
the ones which are definable in $\langle M, < \rangle$.  This
assumption implies that definable subsets of $M^n$ (in the sense of
${\mathcal M}$) admit very nice structure theorems (like the \emph{cell
  decomposition}~\cite{KPS86}) or Theorem~{\ref{uf}} below.  The
following are examples of o-minimal structures: the ordered group of
rationals $\langle\IQ,<,+,0,1\rangle$, the ordered field of reals
$\langle\IR,<,+,\cdot,0,1\rangle$, the field of reals with exponential
function, the field of reals expanded by restricted pfaffian functions
and the exponential function, and many more interesting structures
(see~\cite{vandendries98,Wi96}). An example of non o-minimal structure
is given by $\langle\IR,<,\sin,0\rangle$, since the definable set $\{x
\mid \sin(x)=0\}$ is not a finite union of points and open intervals.
However, let us mention that the
structure\footnote{$\sin_{|_{[0,2\pi]}}$ and $\cos_{|_{[0,2\pi]}}$
  correspond to the sinus and cosinus functions restricted to the
  segment $[0,2\pi]$.} $\langle \IR,+,\cdot,0,1,<,\sin_{|_{[0,2\pi]}},
\cos_{|_{[0,2\pi]}} \rangle$ is o-minimal (see~\cite{vandendries96}).

\begin{thm}[Uniform Finiteness~\cite{KPS86}]\label{uf}%
  Let ${\mathcal M} = \langle M, < , \ldots \rangle$ be an o-minimal
  structure.  Let $S \subseteq M^m \times M^n$ be definable (in $\mathcal
  M$), we denote by $S_a$ the fiber $\{y \in M^n | (a,y) \in S
  \}$. Then there is a number $N_S \in \IN$ such that for each $a \in
  M^m$ the set $S_a \subseteq M^n$ has at most $N_S$ definably
  connected components.
\end{thm}

\subsection{Generalities on O-Minimal Games}

\begin{defi}
  Given $\A$ an $\M$-game, we say that $\A$ is an \emph{o-minimal
    game} if the structure $\M$ is o-minimal and if all transitions
  $(q,g,a,R,q')$ of ${\mathcal A}$ belong to\footnote{This is a particular
    case of reset for $\M$-game where we consider only constant
    functions for resets.}  $Q \times 2^{V_2} \times \Sigma \times
  2^{V_2} \times Q$.
\end{defi}

\label{PA}
Let us notice that the previous definition implies that given $\A$ an
o-minimal game, the guards, the resets and the dynamics are definable
in the underlying o-minimal structure.  We denote by $\P_\A$ the
coarsest partition of the state space $S =Q \times V_2$ which respects
$\Goal$, and all guards and resets in $\A$. Note that $\P_\A$ is a
finite definable partition of $S$.

Due to the strong reset condition we have that $\P_\A$ is stable under
the action of $\Pred_a$ for every action $a$. This holds by the same
argument that allows to decouple the continuous and discrete
components of a hybrid system in~\cite{LPS00}. Let us also notice
that, in the framework of o-minimal games, any refinement of $\P_\A$
is stable under the action of $\Pred_a$ for every $a \in \Sigma$.

\begin{exa}
  The continuous dynamics of timed automata (see Example~\ref{ex-ta})
  is definable in the o-minimal structure $\langle
  \IR,+,0,1,<\rangle$.  The continuous dynamics of rectangular
  automata (see Example~\ref{ex-rect}) is definable in the o-minimal
  structure $\langle \IR,+,\cdot,0,1,<\rangle$. Hence games on timed
  (resp. rectangular) automata with strong resets are particular cases
  of o-minimal games. The $\mathcal M$-game of Example~\ref{ex:spiraletot}
  is in fact an o-minimal game; indeed one can see that it can be
  defined in the structure $\langle
  \IR,+,\cdot,0,1,<,\sin_{|_{[0,2\pi]}}, \cos_{|_{[0,2\pi]}} \rangle$
  which is o-minimal (see~\cite{vandendries96}).
\end{exa}

\subsection{Solving O-Minimal Games}\label{subsec-solvingomin}
In this subsection, we will see how we can (easily) build a partition
which is stable under the actions of the controllable predecessor
operator. The key ingredients to build this partition will be $(i)$
the strong resets conditions and $(ii)$ the superword abstraction. The
finiteness of the obtained partition will be discussed in
Subsection~\ref{subsec-defina}.

\begin{prob}\label{prop-subwordpart}
  Let $\A$ be an o-minimal game, and $\P_\A$ the partition
  corresponding to its guards and resets. The superword (resp. suffix)
  partition $\Sup(\P_\A)$ (resp. $\Suf(\P_\A)$) is stable under the
  action of $\pi_{\textsf{\upshape partial}}$
  (resp. $\pi_{\textsf{\upshape perfect}}$).
\end{prob}

\proof
  This proposition is not a corollary of
  Proposition~\ref{prop:supwordstable}, as $\Sup(\P_\A)$ \emph{is
    not} superword-stable. However, the proof of
  Proposition~\ref{prop:supwordstable} only relied on the fact that in
  a superword-stable partition, two points in a piece of the partition
  have the same superword abstraction,
  which is precisely what we have in the current case. Hence the
  previous proof can be mimicked, and we do not write all details.  It
  is worth noting also that we do not use all properties of o-minimal
  games, but only the strong reset property, which ensures that the
  partition is stable under $\Pred_a$ for every action $a \in
  \Sigma$. \qed

\subsection{Definability and Finiteness Issues.}\label{subsec-defina}
In the previous subsection, we have proved that, given $\A$ an
o-minimal game, the partition $\Sup(\P_\A)$ (resp. $\Suf(\P_\A)$) is
stable under the action of the controllable predecessor operator under
the partial (resp. perfect) observation framework. We will now show
that this partition is finite. For this we will exploit the finiteness
property of o-minimality and in order to do so, we first need to prove
that our encodings are definable.

\subsubsection{Definability.}
Let $(\M,\gamma)$ be a dynamical system and $\P$ be a finite partition
of $V_2$.  We now would like to show that in the case of
\emph{o-minimal dynamical system} the superword encoding previously
discussed can be done in a \emph{definable} way.  The approach closely
follows the one used in~\cite[Section~12.2]{brihaye06} for the suffix
abstraction (called suffix dynamical type in this paper).  \medskip

Let $(\M,\gamma)$ be an o-minimal dynamical system and $\P$ be a
finite definable partition of $V_2$.  First let us notice that, since
$\P$ is finite and definable, given $S \in 2^{\P}$ one can easily
write a first-order formula $\varphi(y,\tau)$ which is true if and
only if ${\mathcal F}_y(\tau) = S$ (where ${\mathcal F}_y$ is defined
similarly to ${\mathcal F}_x$~--~see page \pageref{def:wordgx}). Let us
give this formula, assuming that $S = \{A_1,\ldots,A_n\}$:
\begin{align*}
  \varphi_S(y,\tau) \ \equiv \ & \exists x_1 \ \exists t_1 \ \cdots \
  \exists x_n \ \exists t_n \ \bigwedge_{i=1,\ldots,n} \big(
  \gamma(x_i,t_i)=y ~\wedge~
  \gamma(x_i,t_i+\tau) \in A_i\big)\\
  & \quad \wedge \ \forall x \ \forall t \
  \big(\gamma(x,t)=y \big) \Rightarrow \big(\gamma(x,t+\tau) \in A_1
  \cup \cdots \cup A_n\big).
\end{align*}
Thus, for each $y \in V_2$, the set ${\mathcal F}_y$ exactly consists of
the connected components of the sets $\{ \tau \in M^+ \mid
\varphi_S(y,\tau) \}$, for $S \in 2^{\P}$; \textit{i.e.} ${\mathcal F}_y$
is a set of intervals. In order to show that ${\mathcal F}_y$ is
first-order definable we need to encode each interval $I \subseteq M$
as a point in some cartesian power of $M$. An interval $I \subseteq M$
is entirely characterized by \emph{(i)} its end-points and \emph{(ii)}
the fact of being right (resp. left) open or closed. For \emph{(i)} we
formally need a couple to represent a single end point in order to
recover $-\infty$ and $+\infty$ (as in the projective line case).  For
\emph{(ii)} we can use a binary encoding, let us say $0$ means open
and $1$ closed. Thus any interval $I \subseteq M$ will be encoded by
an element $(a_1,a_2,a_3,b_1,b_2,b_3) \in M^6$. For instance, the
interval $I = \{ x \in \IR \mid x \ge 5\} $ is encoded by
$(5,1,1,1,0,0)$. Thanks to this ``trick'', one can find a first-order
formula $\varphi_y$ defining ${\mathcal F}_y$. The writing of the formula
$\varphi_y$ is not difficult but rather tedious: different cases have
to be considered (depending on whether the interval $I$, encoded by an
element of $M^6$, is left (resp. right) bounded and left (resp. right)
open or closed). Further details of the construction of the formula
can be found in~\cite[Section~12.2]{brihaye06}.

\subsubsection{Finiteness.}
We will now prove that when considering o-minimal dynamical systems,
only finitely many finite superwords are needed to encode all possible
trajectories.

\begin{prob}
  Let $({\mathcal M},\gamma)$ be an o-minimal dynamical system and $\P$ be
  a finite definable partition of $V_2$. There exists finitely many
  finite superwords associated with $(\M,\gamma)$ w.r.t. $\P$.
\end{prob}

\proof
  Given $S \in 2^{\P}$ let us first consider the set
  $${\mathcal F}_y(S) = \big\{ \tau \in M^+ \mid {\mathcal F}_y(\tau)
  = S\big\} = \big\{ \tau \in M^+ \mid \varphi_S(y,\tau) \big\}.$$ By
  the above discussion, the set ${\mathcal F}_y(S)$ is a definable subset
  of $M$. Hence by o-minimality it is a finite union of points and
  open intervals, in particular, it has only finitely many connected
  components. By definition of ${\mathcal F}_y$ we have the following
  equality.
  \[
  |{\mathcal F}_y| = \sum_{S \in 2^{\P}} \Big(\text{number of connected
    components of } {\mathcal F}_y(S)\Big).
  \]
  Since $\P$ is finite we can conclude that ${\mathcal F}_y$ is finite.

  \medskip Using the uniform finiteness theorem (Theorem~\ref{uf}) we
  obtain that there exists $N \in \IN $ such that for all $y \in V_2$
  we have that $\bigl|{\mathcal F}_y \bigr| \le N$.

  In terms of word encoding, this means that there are only finitely
  many superwords associated with the points of the (output) space
  $V_2$.  More precisely, the superwords $\Sup_{\P}(y)$ have lengths
  uniformly bounded by $N$. Since the superwords $\Sup_{\P}(y)$ are
  words on the \emph{finite} alphabet $2^{\P}$, this completes the
  proof.  \qed

The previous proposition directly implies the finiteness of the
partition $\Sup(\P)$. Moreover we have that this partition is
definable, as stated in the following proposition.

\begin{prob}\label{tpfinidefinable}
  Let $({\M},\gamma)$ be an o-minimal dynamical system, $\P$ be a
  finite definable partition of the output space $V_2$. The partition
  $\Sup(\P)$ is finite and definable.
\end{prob} 

\proof
  Since there are only finitely many superwords, it suffices to show
  that given $y \in V_2$ and $SW$ a superword on $\P$ (i.e. a word on
  $2^{\P}$), we can define (by a first-order formula) that
  $SW=\Sup_{\P}(y)$. Suppose that $SW= S_{1} \cdots {S}_{k} \cdots
  S_{n}$, where $S_{k} \in 2^{\P}$. We have that $SW=\Sup_{\P}(y)$ if
  and only if the following formula holds.
  \begin{align*}
    \exists \tau_1 \in M^+, \ \exists \tau_2 \in M^+, \ \cdots \
    \exists \tau_n \in M^+, \ \exists I_1 \in {\mathcal F}_{y}, \ I_2 \in
    {\mathcal F}_{y}, \ \cdots \ \exists I_n \in {\mathcal F}_{y}\\
    (\tau_1 < \tau_2 < \cdots < \tau_n) \ \wedge \
    \bigwedge_{k=1}^n {\mathcal F}_y(\tau_k)=S_k \ \wedge \
     {\mathcal F}_y = \{I_1,I_2,\ldots,I_n\}.
  \end{align*}
  Notice that the above formula is first-order since ${\mathcal F}_y$ is
  first-order definable and testing whether ${\mathcal F}_y(\tau_k)=S_k$
  is also first-order definable.  \qed

\subsection{Synthesis of Winning Strategies}
We now prove that given $\A$ an o-minimal game definable in $\mathcal M$,
we can construct a \emph{definable} strategy (in the same structure
$\mathcal M$) for the winning states under partial observation. The
effectiveness of this construction will be discussed later.

\begin{thm}\label{synthesis}
  Given $\A$ an o-minimal game, there exists a definable memoryless
  winning strategy under partial (resp. perfect) observation for each
  $(q,y) \in \pi_{\textsf{\upshape partial}}^*(\Goal)$
  (resp. $\pi_{\textsf{\upshape perfect}}^*(\Goal)$).
\end{thm}

\proof
  By Proposition~\ref{prop-subwordpart}, the partition $\Sup(\P_\A)$
  is finite, definable and stable under $\pi_{\textsf{\upshape
      partial}}$. In particular, there exists thus $n \in \IN$ such
  that $\pi_{\textsf{\upshape partial}}^*(\Goal) =
  \pi_{\textsf{\upshape partial}}^n(\Goal)$. Hence, by
  Proposition~\ref{proppi2}, $\pi_{\textsf{\upshape
      partial}}^n(\Goal)$ is the set of winning states.

  Given $(q,y) \in \pi_{\textsf{\upshape partial}}^n(\Goal)$, we know
  that there exists a winning strategy from $(q,y)$.  We now have to
  point out a definable winning strategy from $(q,y)$. Following the
  proof of Proposition~\ref{proppi2}, we build the definable strategy
  by induction on the number of iterations of $\pi_{\textsf{\upshape
      partial}}$.  Let us suppose we have already built a strategy on
  each piece of $W = \displaystyle \bigcup_{0 \le i \le
    k}\pi_{\textsf{\upshape partial}}^i(\Goal)$, let us now consider
  $\pi_{\textsf{\upshape partial}}(W) \setminus W$.

  By Proposition~\ref{prop-subwordpart}, we know that
  $\pi_{\textsf{\upshape partial}}(W) \setminus W$ is a finite union
  of pieces of $\Sup(\P_\A)$. Let $P$ be one of these pieces.  We know
  that $P$ corresponds to a finite superword on $\P_\A$. Thus given
  $(q,y) \in P$ we have that
  $$\Sup_{\P_{\A}}(y) = S_1S_2\cdots S_k, \quad 
  \text{ where } S_i \in 2^{\P_{\A}}.$$

  Since $(q,y) \in \pi_{\textsf{partial}}(W) \setminus W$, the
  superword $\Sup_{\P_{\A}}(y)$ contains a prefix $S_1 \cdots S_l$
  such that there is $a \in \Sigma_c$ with:
  \begin{itemize}
  \item for all $P_i \in S_l$,  $P_i \subseteq \Pred_a(W)$,
  \item for all $j \le l$, for all $P_i \in S_j$,
    $\uPred(\overline{W}) \cap P_i = \emptyset$.
  \end{itemize}
  Since for all $P_i \in S_l$, we have that $P_i \subseteq
  \Pred_a(W)$, the controllable action $a \in \Sigma_c$ is such that
  given any $(q,y) \in S_l$ a transition labelled by $a$ is enabled
  and all such transitions lead to $W$. The strategy for $(q,y)$ will
  be to perform action $a$ after some delay. We now explain how to
  choose this delay.

  Let $(q,y)$ be such that $(q,y) \in P$. Let us consider $\Time(y)$
  the subset of $M^+$ defined as follows:
 $$\Time(y)=\{\tau \in M^+ \mid
 \exists y' \in S_l\text{ such that } (q,y) \xrightarrow{\tau}
 (q,y')\}.$$ This set is definable since $S_l$ is definable.
  
 By o-minimality, we have that $\Time(y)$ is a finite union of points
 and open intervals. Let us denote by $I$ the leftmost point or
 interval.  Let us notice that $I$ is definable. If $I$ has a minimum
 $m$, we define $\lambda(q,y)=(m,c)$. Otherwise two cases may
 occur. If $I$ is bounded then it is of the form $(m,m')$ or $(m,m']$
 in this case we define\footnote{Let us recall that every o-minimal
   ordered group is torsion free and divisible (see~\cite{PS86}), this
   implies there exists a unique $y$ satisfying $y+y =(m+m')$, which
   we note $\frac{1}{2}(m+m')$.}
 $\lambda(q,y)=(\frac{1}{2}(m+m'),c)$.  Finally if $I$ has no minimum
 and is unbounded it is of the form $(m,\infty)$ and in this case we
 define $\lambda(q,y)=(m+1,c)$. We summarize\footnote{Let us notice
   that the way we extract a single point from $\Time(y)$ is nothing
   more than the \emph{curve selection} for o-minimal expansions of
   ordered abelian groups, see~\cite[chap.6]{vandendries98}.} the
 definition of $\lambda$ on $S_l$ as follows:
  \[
   \lambda(q,y) = 
   \begin{cases}
     \big(\min(I),c\big) & \text{ if } \ \varphi_1(y)\\
     \big(\frac{1}{2}\big(\inf(I)+\sup(I)\big),c\big) &
     \text{ if } \ \varphi_2(y)\\
     \big(\inf(I) +1,c\big) & \text{otherwise}
   \end{cases}
   \]
   where $\varphi_1(y)$ is a formula which is true if and only if $I$
   (or $\Time(y)$) has a minimum and $\varphi_2(y)$ is a formula which
   is true if and only if $I$ has no minimum and is bounded. Thus
   clearly $\lambda$ is definable.
  
   Since there are finitely many $P \in \Sup({\P_{\A}})$, we can
   conclude that $\lambda$ is definable.  \qed

\begin{rem}
  Note that the memoryless strategy given by Theorem~\ref{synthesis}
  is computable if $\pi_{\textsf{\upshape partial}}^*(\Goal)$ is.
\end{rem}

\begin{rem}
  Let us notice that in the case of timed automata dynamics (described
  in Example~\ref{ex:AT}), our definable strategies correspond to the
  realizable strategies computed in~\cite{BCFL04}.
\end{rem}

\subsection{Decidability Result}\label{subdecid}
Theorem~\ref{synthesis} is an existential result. It claims that given
an o-minimal game, there exists a definable memoryless strategy for
each $y \in \pi_{\textsf{\upshape partial}}^*(\Goal)$, and by
Theorem~\ref{prop-subwordpart} we know that $\Sup(\P_{\A})$ is finite.
The conclusion of the previous subsection is that given an o-minimal
game there exists a definable memoryless winning strategy for each $y
\in \pi_{\textsf{\upshape partial}}^*(\Goal)$.

In general, Theorem~\ref{synthesis} does not allow to conclude that
the control problem in an $\M$-structure is decidable. Indeed it
depends on the decidability of $\Th(\M)$. We can state the following
theorem:

\begin{thm}\label{decidability}
  Let $\M$ be an o-minimal structure such that $\Th(\M)$ is decidable
  and $\C$ a class of $\M$-automata. Then the control problem under
  partial (resp. perfect) observation in class $\C$ is
  decidable. Moreover if $\A \in \C$, the set of winning states
  $\pi_{\textsf{\upshape partial}}^*(\Goal)$
  (resp. $\pi_{\textsf{\upshape perfect}}^*(\Goal)$) under partial
  (resp. perfect) observation is computable and a memoryless winning
  strategy can be effectively computed for each $(q,y) \in
  \pi_{\textsf{\upshape partial}}^*(\Goal)$
  (resp. $\pi_{\textsf{\upshape perfect}}^*(\Goal)$).
\end{thm}

\proof
  By Proposition~\ref{tpfinidefinable}, for each $\A \in \C$, $\Sup
  (\P_\A)$ is a definable finite partition respecting
  $\Goal$. Moreover by Proposition~\ref{prop-subwordpart}, $\Sup
  (\P_\A)$ is stable under $\pi_{\textsf{\upshape
      partial}}$. Hypothesis of Corollary~\ref{cor:etatsgagnants} are
  thus satisfied and we get that the control problem in class $\C$ is
  decidable and that the winning states of a game $\A \in \C$ are
  computable. Moreover Theorem~\ref{synthesis} ensures that a
  memoryless strategy can be effectively defined from such winning
  states.  \qed

\begin{rem}
  $\langle \IR,<,+,0,1\rangle$ and $\langle \IR,<,+,\cdot,0,1\rangle$
  are examples of o-minimal structures with decidable theory and so
  o-minimal games based on theses structures can be solved by
  Theorem~\ref{decidability}.
\end{rem} 

\begin{rem}
In this paper we did not distinguish Zeno behaviours. In particular,
in our framework, if the environment has a strategy that prevents the
game to reach the $\Goal$ locations by blocking time, we say that the
controller loses the game. In the framework of timed automata, an
\textit{ad-hoc} solution to this \emph{problem of Zenoness} has been
proposed in~\cite{AFH+03}. However, due to the strong reset conditions
of o-minimal hybrid systems, the method of~\cite{AFH+03} cannot be
easily applied to our framework, but this problem is somehow
orthogonal to ours.
\end{rem}

\section{Conclusion}

In this paper we have studied games based on dynamical systems with
general dynamics, both under a prefect and a partial observation of
the dynamics. Under the first hypothesis, we have shown that
time-abstract bisimulation is not fine enough to solve these games,
which is a major difference with the case of timed automata. By means
of an encoding of trajectories by words, we have obtained a good
abstraction for control problems (with reachability winning
conditions, but it applies also to basic safety winning
conditions). We have finally provided decidability and computability
results for o-minimal games under both perfect and partial observation
hypothesis. Our technique applies to timed automata, and we recover
decidability of timed games~\cite{AMPS98}, as well as the construction
of winning strategies~\cite{BCFL04} as side results.

\section*{Acknowledgment}

The two first authors have been partly supported by the ESF project
GASICS.  The first author has been partly supported by the project
DOTS (ANR-06-SETI-003) and by the EU project QUASIMODO.  The second
author has been partly supported by a grant from the National Bank of
Belgium and by a FRFC grant: 2.4530.02. 

\newcommand{\etalchar}[1]{$^{#1}$}

\end{document}